\documentclass[journal,10pt]{IEEEtran}

\makeatletter
\def\ps@headings{%
	\def\@oddhead{\mbox{}\scriptsize\rightmark \hfil \thepage}%
	\def\@evenhead{\scriptsize\thepage \hfil \leftmark\mbox{}}%
	\def\@oddfoot{}%
	\def\@evenfoot{}}
\makeatother 

\usepackage{graphicx}
\usepackage{amsmath, amsfonts,epsfig, multirow, floatflt,float}
\usepackage{amssymb}
\usepackage{cite}
\usepackage{algorithm,algorithmic}
\usepackage{hyperref}
\usepackage{enumitem}
\usepackage{mathrsfs}
\usepackage{url}
\usepackage{color}
\usepackage{subfigure}
\usepackage{array}
\usepackage{makecell}

\allowdisplaybreaks

\graphicspath{{figure/}}

\begin{document}

\title{Multi-Timescale Control and Communications with Deep Reinforcement Learning---Part I: Communication-Aware Vehicle Control}

	\author{Tong Liu, Lei~Lei, {\it Senior Member, IEEE}, Kan~Zheng, {\it Senior Member, IEEE}, Xuemin (Sherman) Shen, {\it Fellow, IEEE}}

\maketitle

{\begin{abstract}
An intelligent decision-making system enabled by Vehicle-to-Everything (V2X) communications is essential to achieve safe and efficient autonomous driving (AD), where two types of decisions have to be made at different timescales, i.e., vehicle control and radio resource allocation (RRA) decisions. The interplay between RRA and vehicle control necessitates their collaborative design. In this { two-part} paper (Part I and Part II), taking platoon control (PC) as an example use case, we propose a joint optimization framework of multi-timescale control and communications (MTCC) based on Deep Reinforcement Learning (DRL). In { this paper (Part I)}, we first decompose the problem into a communication-aware DRL-based PC sub-problem and a control-aware DRL-based RRA sub-problem. Then, we focus on the PC sub-problem assuming an RRA policy is given, and propose the MTCC-PC algorithm to learn an efficient PC policy. To improve the PC performance under random observation delay, the PC state space is augmented with the observation delay and PC action history. Moreover, the reward function with respect to the augmented state is defined to construct an augmented state Markov Decision Process (MDP). It is proved that the optimal policy for the augmented state MDP is optimal for the original PC problem with observation delay. Different from most existing works on communication-aware control, the MTCC-PC algorithm is trained in a delayed environment generated by the fine-grained embedded simulation of C-V2X communications rather than by a simple stochastic delay model. Finally, experiments are performed to compare the performance of MTCC-PC with those of the baseline DRL algorithms.
\end{abstract}}


\section{Introduction}

The Cellular Vehicle-to-Everything (C-V2X) system provides message delivery services for vehicular applications using fourth-generation (4G) and fifth-generation (5G) cellular connectivity \cite{gyawali2020challenges}. Due to its ability to provide ubiquitous coverage, high reliability, and low latency, the C-V2X system is crucial for autonomous vehicles (AVs) \cite{9611166}. Meanwhile, AVs are seen as a major driving use case for enhancing C-V2X communications in six-generation (6G) wireless system \cite{yang2021edge}. Designing vehicle control-oriented C-V2X system falls into the category of Networked Control Systems (NCS) research, where closed-loop control relies on data transmission in communication networks\cite{zhang2019networked}. \par

In contrast to the conventional network design, the performance of NCS is measured in terms of the efficiency in accomplishing a control task rather than the network performance metrics such as throughput and delay\cite{redder2019deep,ayan2022task}. Compared with pure NCS, the ecosystem of C-V2X is more complex due to the co-existence of safety-critical vehicle control applications as well as non-safety applications such as infotainment. Since the latter type of applications usually  {requires} high throughput, an effective radio resource allocation (RRA) mechanism is indispensable for assigning the limited network resources to various applications, guaranteeing the safety and efficiency of vehicle control tasks while maximizing the throughput of non-safety applications. For this purpose, RRA in C-V2X systems should be control-aware, taking into account the control performance degradation due to the delay or packet loss in control-related information delivery.  \par

Meanwhile, the amount of control performance degradation heavily depends on the robustness of vehicle controllers to non-ideal communications. Conventional controllers of AVs are usually designed based on control theory under the assumption of zero-delay and zero-loss communications\cite{zheng2015stability,li2017dynamical}. In order to reduce the effects of communication impairments on control performance, vehicle controllers should be communication-aware, considering the statistical properties of random delay and packet loss in C-V2X communications. \par 
\subsection{ {Collaborative Design of Communications and Control}}
The interplay between RRA and vehicle control necessitates the collaborative design of communications and control functions. Existing works mainly tackle the problem in two directions, i.e., control-aware communications and communication-aware control. 

\subsubsection{ {Control-aware communications}}
Control-aware or task-oriented communications aim at scheduling network resources to achieve satisfactory control performance. In order to characterize the significance of transmitted information in achieving the control target, two cross-layer performance metrics are often adopted to guide the optimization. The most widely used metric is Age of Information (AoI)\cite{kaul2012real}, which captures the importance of information by measuring its timeliness attribute. Meanwhile, another metric, i.e., Value of Information (VoI), measures how much the recipient of the information can reduce the uncertainty of the stochastic processes related to decision-making \cite{kosta2017age}. Since the co-design problem in this research direction is tackled from communications perspective, the considered controllers are normally quite simple and are designed based on conventional control theory and ideal communications assumption.\par 

\subsubsection{ {Communication-aware control}}
On the other hand, communication-aware or delay-aware control aims at designing controllers that are robust to communication imperfections. Examples are networked control that analyzes the tolerance of controllers to delay and packet loss using mathematical models; and event-triggered control that determines whether or not to sample and transmit system signals based on event or time. Since the co-design problem in this research direction is tackled from control perspective, non-ideal communications are usually modeled as either constant delay or stochastic delay under coarse-grained surrogate communication models that are control-agnostic \cite{li2018nonlinear,huang2022design,xu2022stochastic,xu2019modeling,ma2020distributed,wang2020adaptive,li2019platoon}. \par

\subsection{ {Motivations}}
While most existing works of NCS study the co-design problem from either the communications or control perspective, it is our hypothesis that great benefits will arise from joint optimization in a unified perspective, where both components are designed using advanced technologies and are aware of the necessary details of the other components. Vehicle controllers are conventionally designed based on classical control theory, such as linear controller, $\mathcal{H}_\infty$ controller, and Sliding Mode Controller (SMC), etc.\cite{Li2017,Sinan2014,Yang2021}; while RRA in C-V2X systems is traditionally studied using optimization theory. One of the main limitations of such approaches is that rigorous mathematical models are required, which are either inaccurate or unavailable for real-world problems; or it is computationally expensive to solve the models. Meanwhile, both vehicle control and RRA are Sequential Stochastic Decision Problem (SSDP), where a sequence of decisions have to be made over a specific time horizon for a dynamic system whose states evolve in the face of uncertainty. As a promising approach to solve SSDP, Deep Reinforcement Learning (DRL) has attracted considerable attention in recent years and has been adopted for vehicle control and RRA as an emerging trend. DRL inherits the model-free learning
capability from Reinforcement Learning (RL), which can learn an optimal control policy directly from experience data by trial and error without knowledge of the underlying SSDP model. Moreover, it deals with the curse-of-dimensionality problem of RL by approximating the value functions and/or policy functions using deep neural networks (DNN) \cite{hinton2006reducing,lei2020deep}. We believe that \textbf{tackling both vehicle control and RRA problems under a unified DRL framework will better reveal the inter-dependency between the two components} and thus facilitate the joint optimization task. \par   

Since the frequency of vehicle control and sampling (normally between $0.01$ second (s) to $0.1$ s) is often lower than that of RRA (e.g., $1$ millisecond (ms) in C-V2X), \textbf{the joint optimization of RRA and vehicle control is generally a multi-timescale decision problem.} The most straightforward approach is solving an integrated full-space model containing detailed vehicle control and RRA sub-models. However, simultaneous derivation of vehicle control and RRA decisions at multi-timescales yields a large-scale optimization problem, which is computationally infeasible even for modern machine learning techniques. Our main goal in this {  two-part} paper (Part I and Part II) is to propose an efficient DRL-based approach for multi-timescale control and communications (MTCC) in the C-V2X system. To the best of our knowledge, \textbf{this is the first paper that jointly optimizes multi-timescale vehicle control and RRA decisions under a unified DRL framework.}\par

As there are a variety of control tasks for AVs, \textbf{we will focus on platoon control (PC) as an example use case.} Meanwhile, the modular nature of the proposed approach enables its extension to other AV tasks. As a basic function of AVs, PC aims to determine the control inputs for following vehicles so that all vehicles move at the same speed while maintaining the desired distance between each pair of preceding and following vehicles \cite{li2017dynamical}. Although PC can be performed without information exchange between vehicles based on the adaptive cruise control (ACC) functionality, the more advanced cooperative
adaptive cruise control (CACC) extends ACC with V2X communications and is capable of improving the PC performance by reducing the inter-vehicle distance while guaranteeing string stability\cite{Dey2016}. \par 

\subsection{ {Contributions}}
The main contributions of this {  two-part} paper (Part I and Part II) are explained below.
\begin{itemize}
	\item \textbf{Unified DRL framework for multi-timescale control and communications:}
	The time horizon is divided into control intervals, where each control interval consists of multiple communication intervals. Instead of employing the full-space approach with formidable computation complexity, we decompose the problem into two sub-problems, i.e., (1) communication-aware DRL-based PC, and (2) control-aware DRL-based RRA. We propose the MTCC-PC algorithm to learn the PC policy assuming an RRA policy is given, and the MTCC-RRA algorithm to learn the RRA policy assuming a PC policy is given. Finally, \emph{A sample- and computational-efficient approach is proposed to jointly learn the PC and RRA policies by training MTCC-PC and MTCC-RRA algorithms in an iterative process.}

	\item\textbf{Integrated DRL model capturing the interplay between RRA and PC:}	
	We conceive a communication-aware DRL model for PC and a control-aware DRL model for RRA, both of which are integrated parts of the multi-timescale decision framework. Specifically, we \emph{augment the PC state space with the observation delay}, which serves as a bridge between the PC and RRA models. Moreover, we \emph{incorporate the advantage function of the PC model in the RRA reward function, which quantifies the amount of PC performance degradation caused by observation delay}. Finally, we \emph{augment the state space of RRA with control input history} for a more well-informed RRA policy. Since both PC and RRA are formulated into DRL models, it is much easier to fully consider the interplay between them. Specifically, the MTCC-PC algorithm is trained in a delayed environment generated by the fine-grained embedded simulation of C-V2X communications rather than by a simple stochastic delay model. Moreover, the RRA decisions in the MTCC-RRA algorithm are made based on the ``VoI per control interval", which provides a finer-grained VoI compared with the existing VoI calculation methods.

	\item\textbf{Efficient DRL solution addressing random observation delay, multi-agent and sparse reward problems:}
	To improve the performance of PC in the face of random observation delay, we \emph{augment the PC state space with PC action history and prove the Markov property of the augmented state}. Moreover, we \emph{define the reward function for the augmented state to construct an augmented state Markov Decision Process (MDP), and prove that the optimal policy for this MDP is also optimal for the original PC problem with observation delay}. To deal with the multi-agent problem in the MTCC-RRA algorithm, we apply the \emph{reward shaping technique to design an individual reward for each agent}, so that they can deduce their own contributions in the global reward. Moreover, to tackle the sparse reward problem in RRA, we use \emph{reward backpropagation prioritized experience replay (RBPER)} technique\cite{zhong2017reward} to improve training efficiency. \par
	
\end{itemize}
\begin{figure}[tb!]
	\centering
	\includegraphics[width=0.45\textwidth]{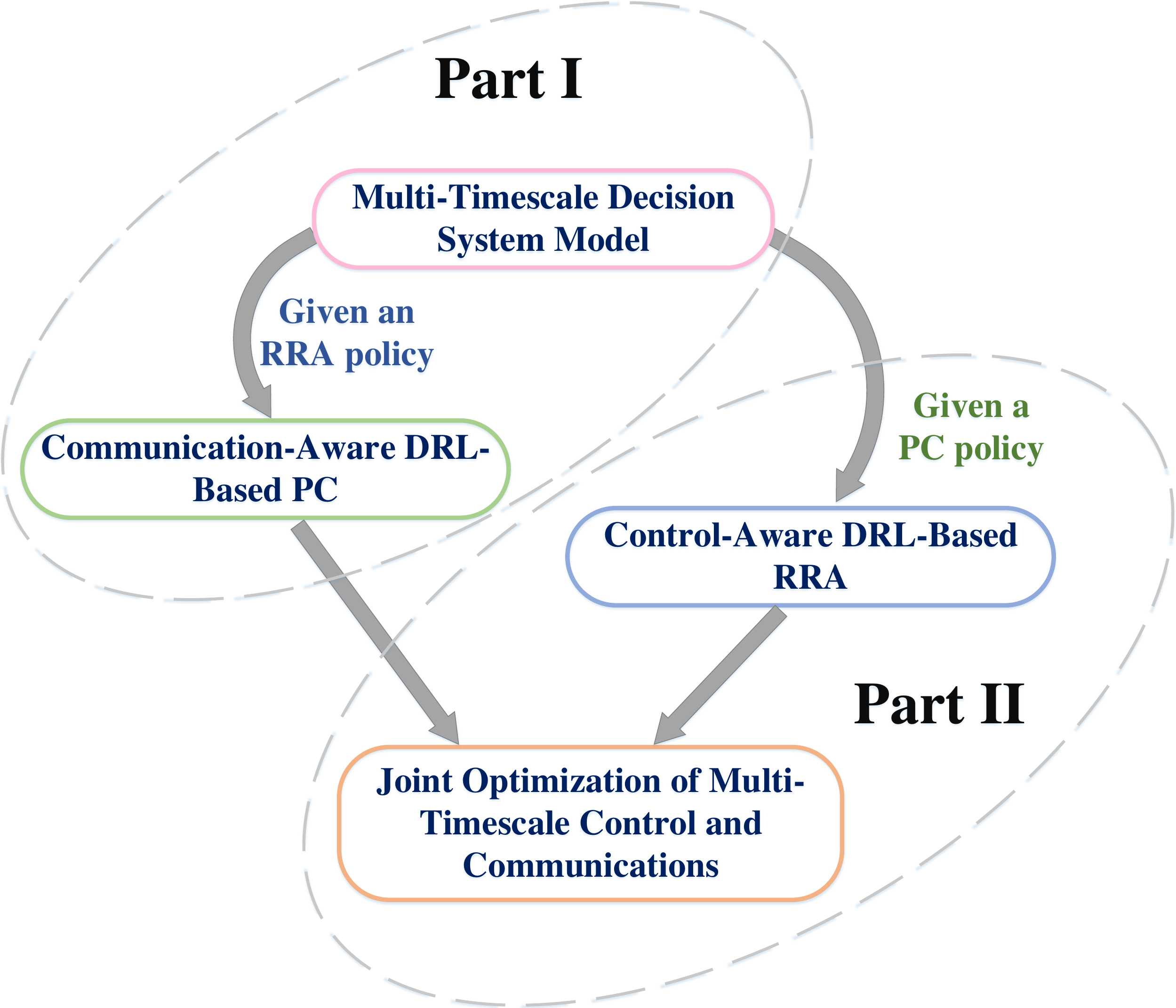}
	\caption{The organization of the two-part paper}
	\label{contents}
\end{figure}

\subsection{ {Organization of the Two-Part Paper}}
The organization of this { two-part} paper is shown in Fig.\ref{contents}. First, the multi-timescale decision system model is introduced in Part I. Then, in order to jointly optimize the multi-timescale PC and RRA decisions using DRL, we assume that an RRA policy is given and study the communication-aware DRL-based PC in Part I. Next, we assume that a PC policy is available and focus on the control-aware DRL-based RRA in Part II. Finally, the joint learning approach that iteratively learns the PC and RRA policies is provided in Part II. \par

The rest of { this paper (Part I)} is organized as follows. Section II introduces the related work. The MTCC system model is presented in Section III. Subsequently, Sections IV and V introduce the communication-aware DRL model for PC and the corresponding DRL solution, i.e., MTCC-PC algorithm, respectively. Then, section VI conducts experiments to demonstrate the effectiveness of the proposed algorithm. Finally, Section VII concludes the paper.\par

\section{Related Work}
\subsection{RRA in C-V2X systems}
Existing works on RRA in C-V2X systems can be categorized into traditional methods and DRL methods. While traditional methods seek solutions to RRA based on classical optimization theory \cite{li2019joint,jameel2020efficient,lien2018low,abanto2018impact,mei2018joint,han2022longitudinal}, DRL methods have gained increasing attention in RRA research \cite{omidshafiei2017deep,ye2019deep,yang2019intelligent,zia2019distributed,nan2021delay,liang2019spectrum,vu2020multi,xiang2021multi,zhang2022mean,parvini2023aoi,xu2023deep} due to its success in learning decision-making policies in a variety of fields recently. There are two typical working modes in C-V2X, named Vehicle-to-Infrastructure (V2I) and Vehicle-to-Vehicle (V2V). It is generally considered that V2I links are mainly used to carry high bandwidth content, while V2V links are mostly used to deliver safety-critical messages{\cite{3GPPr14v2x,3GPPr15v2x,molina2017lte,liang2019spectrum,xiang2021multi,vu2020multi}}. Therefore, the objective of RRA in most existing works is to efficiently share the frequency spectrum between V2I and V2V links, striking a trade-off between maximizing V2I throughput and minimizing V2V delay. \par

Due to the scalability issue of the centralized solution, many works formulate RRA as multi-agent RL (MARL) problems where each V2V agent can only observe its own local state. To solve MARL, the independent learner (IL) approach that directly applies single-agent RL algorithms has been adopted by a few works \cite{omidshafiei2017deep,ye2019deep,yang2019intelligent,zia2019distributed,nan2021delay}. While IL is simple and scalable, the concurrent exploration of multiple agents can lead to non-stationarity issues, especially for DRL algorithms with experience replay. 
To deal with this challenge, the fingerprint-based technique is adopted in \cite{liang2019spectrum} and hysteretic Q-learning and concurrent experience replay trajectory approaches are leveraged in \cite{xiang2021multi}. In \cite{zhang2022mean}, the mean-field game theory is used to enhance scalability and reduce the complexity of MARL solutions. However, another important issue in multi-agent RL namely the credit assignment problem, is not considered in the above works.\par

While most RRA algorithms in C-V2X are agnostic to vehicle control tasks, a few algorithms are specifically designed to support PC applications. The fingerprint-based Deep Q Networks (DQN) algorithm similar to \cite{liang2019spectrum} is used in \cite{vu2020multi} for platoon-based C-V2X system. AoI-aware RRA algorithms are proposed in \cite{parvini2023aoi} based on the multi-agent deep deterministic policy gradient (MADDPG) algorithm. \cite{xu2023deep} formulated a multi-objective RRA problem, which is divided into a set of scalar optimization sub-problems that are modeled as the partially observable stochastic game (P-OSG) and solved by the dual-clip proximal policy optimization (CD-PPO) algorithm. By modeling the network traffic based on message delivery characteristics of PC, the above works have designed efficient RRA algorithms that can better support PC applications. However, the optimization objectives are either delay, AoI, or transmission success ratio, which fail to capture the extent to which the PC performance will be degraded by receiving stale information.\par

\subsection{Delay-aware PC and DRL-based PC}
The existing works on delay-aware PC mainly focus on designing platoon controllers that are robust to communication delay, and deriving the upper bound of communication delay satisfying the internal and string stability for the controllers \cite{li2018nonlinear,huang2022design,xu2022stochastic,xu2019modeling,ma2020distributed,wang2020adaptive}. These works have achieved impressive results on improving PC tolerance to communication delay. However, the delay models are abstract and simple, which cannot accurately reflect the delay distribution induced by the advanced RRA mechanisms in C-V2X communications. Moreover, the platoon controllers are designed based on classical control theory, and thus have limited capability in dealing with the uncertainty and randomness of the environment. \par 

In order to better cope with the uncertain driving environment, nonlinear vehicle dynamics, and real-time application requirement, some recent works study DRL-based PC \cite{wang2018novel,Chu2019,Yan2021,9951132}, where Deep Deterministic Policy Gradient (DDPG) is the most widely used algorithm. However, these works do not consider communication delay when the vehicles share the driving information.\par

In the field of theoretical research on RL, there have been some works on how the agents make decisions when delays occur in one or more forms including observation delay, action delay, and reward delay. The pioneering work of \cite{altman1992closed} considers constant delay scenario, and reformulates the decision process with delays into an augmented state MDP without delays, where the action history is included in the augmented state. The history horizon is from the time step when the delayed observation was generated to one time step before the current state. However, since the assumption of constant delay is usually unrealistic, recent works in this area mostly focus on the random delay scenario. \par


To solve the problem of uncertain augmented state dimension when the random delay occurs, \cite{katsikopoulos2003markov} and \cite{nath2021revisiting} assume that the MDP freezes from the perspective of the agent, i.e., the agent does not take any new actions till the most recent state becomes observable. This may not be possible in practice when the agent must take new actions to interact with the environment at each time step. To deal with random reward delay, the authors augment the state with the time step at which the last observed delayed state was first observed. Moreover, they show that action delay and observation delay are equivalent in the sense that their respective decision processes with delay are both reducible to the MDP with augmented state. In \cite{chen2023efficient}, the delayed time steps after observing the last delayed state is included in the augmented state. However, the authors do not mention how to solve the problem of the uncertain augmented state dimension. Moreover, it is assumed that multiple data packets cannot arrive simultaneously at the same time step, which is not the case in C-V2X communication. Different from the above works, \cite{bouteiller2020reinforcement} augments the state with the action history from maximum delay to the previous time step, which solves the uncertain augmented state dimension problem. The observation delay at each time step is also included in the augmented state. In our work, the augmented state is in a similar form to that defined in \cite{bouteiller2020reinforcement}. Moreover, we provide rigorous proofs of the Markov property of the augmented state, and of the functional equivalence of the augmented state MDP with the original decision process with delay, which are lacking in \cite{bouteiller2020reinforcement}.  \par 


\subsection{Joint optimization of RRA and PC}
While the RRA and PC problems are studied separately in the above works, some recent literature tackles the co-design of them. Most works fall into the class of control-aware communications or communication-aware control, emphasizing the novel design of either communications or control while considering the requirements or constraints posed by existing control or communications mechanisms.    \par

For control-aware communications, \cite{wen2018cooperative} derived sufficient conditions to meet platoon stability of a sampled-data feedback controller, which is used for parameter design of event-triggered communication mechanisms. In \cite{han2022longitudinal}, the communication delay constraints that guarantee plant
stability and string stability are first derived for a non-linear controller, and the obtained delay constraints are used to guide RRA. \par

For communication-aware control, a nonlinear consensus-based platoon controller was proposed in \cite{li2019platoon} based on the probability of successful communication inferred from the carrier sense multiple access with collision avoidance (CSMA/CA) mechanism. \cite{zeng2019joint} first derives an approximate expression for the probability that the wireless system meets the control system\textquoteright s delay needs, and then  {optimizes} the control parameters of the optimal velocity model (OVM) to maximize the probability. In \cite{oliveira2021co}, PC is modeled as a consensus problem, where the parameters of a linear controller are dynamically adjusted according to different information typologies (IFTs) and delays.  \par

Finally, several research works consider the novel design of both PC and communication mechanisms. \cite{hong2020joint} designed a modified distributed model predictive controller (DMPC), which takes into account the set of vehicles whose messages are successfully received. Moreover, a communications approach to select relay vehicles to forward the information of the leading vehicle is proposed, which aims at maximizing the minimal average signal-to-noise ratio (SNR) among the vehicles in the platoon. In \cite{mei2018joint}, the PC and RRA are jointly optimized in order to minimize the tracking error while guaranteeing the minimum SNR requirements of V2V communications and string stability of the platoon. Since the optimization problem is non-deterministic polynomial hard (NP-hard), it is decomposed into separate RRA and PC problems in two stages. The bipartite graph matching method is first used to approximate the subframe allocation scheme, and then the parameters of a linear controller are optimized. Both \cite{hong2020joint} and \cite{mei2018joint} consider non-ideal communications in terms of reliability instead of delay. Instead of focusing on the impact of communication impairments on PC, \cite{zhang2023joint} contemplated the interplay between PC and communications from a different perspective and attends to the effect of PC inputs on the reliability of V2I communications through vehicle mobility behaviors. A joint optimization scheme based on MPC was proposed for RRA and PC, with the aim of maximizing the communication reliability of V2I and minimizing the traffic oscillation of PC.  Our work differs from the above research in considering that the control and communications decisions are usually made at different time scales. Moreover, the co-design is performed under a unified DRL framework, which does not suffer from model inaccuracy or high computational complexity as in conventional control theory, optimization theory, or MPC.\par

\section{System Model}
We consider a platoon with a number of $N>2$ vehicles, i.e., $\mathcal{V}=\{0,1,\cdots, N-1\}$. All vehicles communicate with one another using C-V2X communications. The important symbols used in this paper are summarized in Table \ref{symbol}.\par 
   \begin{table*}[htb!]
			\renewcommand{\arraystretch}{1.2}
		\setlength\tabcolsep{1.8pt}  
		\centering
		\caption{Summary of important symbols used}
		\begin{tabular}{c|c|l}
			\hline
			\textbf{Category} & \textbf{Symbol}&\textbf{Definition} \\
			\hline
			\multirow{16}{*}{\textbf{Control}}&$T$&The control interval\\
           \cline{2-3} 
			&$N$ &The number of vehicles in a platoon \\
           \cline{2-3} 
			&$p_{i,k}$&The one-dimensional position of vehicle i at control interval $k$\\
           \cline{2-3} 
			&$v_{i,k}$&The velocity of vehicle $i$ at control interval $k$\\
           \cline{2-3}  
			& $acc_{i,k}$ & The acceleration f vehicle $i$ at control interval $k$\\
           \cline{2-3} 
			&$e_{pi,k}$& The gap-keeping error of vehicle $i$ at control interval $k$\\
           \cline{2-3} 
           &$e_{vi,k}$& The velocity error of vehicle $i$ at control interval $k$\\
           \cline{2-3} 
           & $j_{i,k}$ & The jerk of vehicle $i$ at control interval $k$\\
           \cline{2-3} 
           & $\tau_{i,k}$ &The observation delay of vehicle $i$ at control interval $k$\\
			\cline{2-3} 
			&$S^{\rm CL}_{i,k}$&The PC state of vehicle $i$ at control interval $k$\\
           \cline{2-3} 
			&$a^{\rm CL}_{i,k}$&The PC action of vehicle $i$ at control interval $k$\\
           \cline{2-3} 
			&$R^{\rm CL}_{i,k}$&The PC reward of vehicle $i$ at control interval $k$\\
           \cline{2-3} 
			&$J^{\rm CL}_{i}$& The expected cumulative reward of PC agent $i$ \\
           \cline{2-3} 
			&$Q^{\rm CL}_{i}$& The Q-value of PC agent $i$ \\
                \cline{2-3} 
                &$V^{\rm CL}_{i}$& The value function of PC agent $i$ \\
                \cline{2-3} 
                \cline{2-3} 
			&$\pi^{\rm CL}_{i}$&The policy of PC agent $i$\\
           \cline{2-3} 
			&$A_{\pi^{\rm CL}_{i}}$&The advantage function of policy $\pi^{\rm CL}_{i}$\\
		 \hline
			\multirow{27}{*}{\textbf{Communication}}
			&$M$ & The number of V2I links\\
           \cline{2-3} 
			&$W$& The bandwidth of sub-channel \\
           \cline{2-3} 
		    &$N_{c}$&The constant CAM size\\
           \cline{2-3} 
		&$N_Q$& The buffer capacity in the number of CAM \\
		\cline{2-3} 
           &$\gamma_{m,(k,t)}$&The SINR of the V2I link $m$ over the sub-channel $m$ at communication interval $(k,t)$ \\
           \cline{2-3} 
           &$\gamma_{i,m,(k,t)}$& The SINR of the V2V link $i$ over the sub-channel $m$ at communication interval $(k,t)$\\
           \cline{2-3} 
            &$I_{i,m,(k,t)}$&The total interference power received by V2V link $i$ over sub-channel $m$\\
            \cline{2-3} 
           &$P^{\rm I}_{m}$&The transmit power of V2I link $m$ over the sub-channel $m$ \\
           \cline{2-3} 
           &$P^{\rm V}_{i,m,(k,t)}$& The transmit power of V2V link $i$ over the sub-channel $m$ at communication interval $(k,t)$\\
           \cline{2-3} 
           &$G_{m,(k,t)}$& The channel gain of the V2I link $m$\\
           \cline{2-3} 
           &$G_{i,B,m,(k,t)}$&  The interference channel gain from V2V link $i$ transmitter to V2I link $m$ receiver\\
           \cline{2-3} 
           &$G_{i,m,(k,t)}$& The channel gain of the V2V link $i$ over the sub-channel $m$\\
           \cline{2-3} 
           &$G_{B,i,m,(k,t)}$& The interference
            channel gain from V2I link $m$ transmitter to V2V link $i$ receiver \\
           \cline{2-3} 
           &$G_{j,i,m,(k,t)}$&The interference channel gain from the V2V link $j$ transmitter to the V2V link $i$ receiver \\
           \cline{2-3} 
           &$\theta_{i,m,(k,t)}$&\makecell[l]{The binary allocation indicator indicating whether V2V link $i$ occupies sub-channel $m$ at communication interval $(k, t)$\\ or not, $\theta_{i,m,(k,t)} \in \{0,1\}$}\\
           \cline{2-3} 
           &$r^{\rm CAM}_{i,(k,t)}$& The transmission rate of V2V link $i$ in terms of CAM at communication interval $(k,t)$\\
           \cline{2-3} 
           &$r_{m,(k,t)}$&The instantaneous data rate of V2I link $m$ at communication interval $(k,t)$\\
            \cline{2-3} 
           &$q^{\rm CAM}_{i,(k,t)}$&The queue length of vehicle $i$ in the number of CAM at communication interval $(k,t)$\\
           \cline{2-3} 
           &$S^{\rm CM}_{i,(k,t)}$&The RRA state of vehicle $i$ at communication interval $(k,t)$\\
           \cline{2-3} 
           &$a^{\rm CM}_{i,(k,t)}$&The RRA action of vehicle $i$ at communication interval $(k,t)$\\
           \cline{2-3} 
           &$R^{\rm CM}_{(k,t)}$&The RRA reward of vehicle $i$ at communication interval $(k,t)$\\
           \cline{2-3} 
           &$R_{{\rm I},(k,t)}$&The RRA reward component related to V2I throughput  at communication interval $(k,t)$\\
           \cline{2-3} 
           &$Q^{\rm CM}_i$&The Q-value of RRA agent $i$\\
           \hline
		\end{tabular}
		\label{symbol}
	\end{table*}

\subsection{Multi-timescale Decision-Making Framework}

\begin{figure}[tb!]
	\centering
	\includegraphics[width=0.47\textwidth]{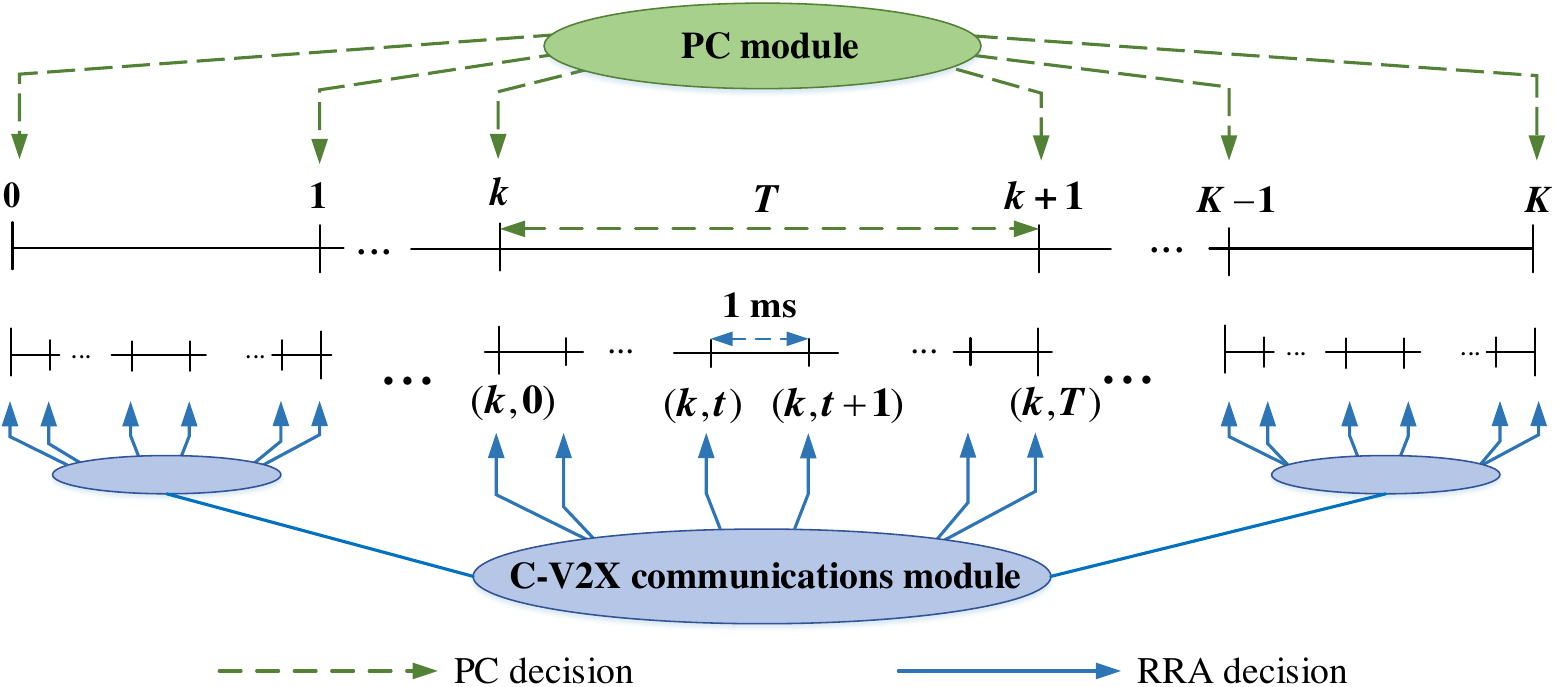}
	\caption{Multi-timescale decision-making framework}
	\label{fig_2}
\end{figure}
As shown in Fig.~\ref{fig_2}, the PC problem is considered within a finite time horizon, which is discretized into $K$ equal-length control intervals indexed by $k \in \mathcal{K} = \{0,1,\cdots, K-1\}$. The duration of each control interval is $T$ milliseconds ($\rm ms$). The vehicle control input (commanded acceleration) $a^{\rm CL}_{i,k}$ for any vehicle $i\in\mathcal{V}$ is applied at time $kT$ and holds constant within time period $[kT,(k+1)T)$. In the rest of the paper, we will use $x_{k}:=x(kT)$ to represent any variable $x$ at control interval $k$. \par

At each control interval $k$, the PC module of each following vehicle (i.e., follower) $i\in\mathcal{V}\backslash \{0\}$ determines the vehicle control input $a^{\rm CL}_{i,k}$ based on the observations of the system state. The vehicle driving status is sampled at time $kT$ (i.e., the sampling period is $T$ $\rm ms$). Specifically, the position $p_{i,k}$, velocity $v_{i,k}$ and acceleration $acc_{i,k}$ of follower $i$ are measured locally by the fusion of inertial navigation and Global Positioning System (GPS). Here $p_{i,k}$ represents the one-dimensional position of the center of the front bumper of vehicle $i$ at control interval $k$. Additionally, each follower $i$ can obtain the driving status $p_{j,k}$, $v_{j,k}$ and $acc_{j,k}$ of the other vehicles $j\in\mathcal{V}\backslash \{i\}$ via V2V communications.  \par


We adopt the Predecessors Following (PF) IFT{ \cite{han2022longitudinal,9951132,ploeg2011design}}, where the Collaborative Adaptive Message (CAM) $c_{i-1,k}=\{p_{i-1,k},v_{i-1,k},acc_{i-1,k}\}$ of the preceding vehicle (i.e., predecessor) $i-1\in\mathcal{V}\backslash \{N-1\}$ are transmitted to the follower $i$. For this purpose, each control interval $k$ is further divided into $T$ communication intervals indexed by $t \in \mathcal{T}=\{0,1,..., T-1\}$ on a faster timescale. Each communication interval has a length of $1$ $\rm ms$ corresponding to the subframe duration in C-V2X communications. The vehicles transmit CAM at time $kT+t, k \in \mathcal{K}, t \in \mathcal{T}$, where the corresponding communication interval is represented as $(k,t)$. Dynamic scheduling is considered, where the C-V2X communication module makes RRA decisions at each communication interval $(k,t)$. In the integrated model, temporal integrity is maintained by $(k,T)=(k+1,0)$. In the rest of the paper, we will use $x_{(k,t)}:=x(kT+t)$ to represent any variable $x$ at communication interval $(k,t)$.\par

Since the PC decisions are made with a coarse time grid of every $T$ $\rm ms$, while the RRA decisions are made with a fine time grid of every $1$ $\rm ms$, we have a multi-timescale decision-making problem. \par

\subsection{Platoon Control Module}

{Each vehicle $i\in\mathcal{V}$ obeys the dynamics model approximated by a first-order system. The state space model in discrete time is derived on the basis of forward Euler discretization: 
	\begin{equation}
	\label{eq2}
	{p}_{i,k+1}={p}_{i,k}+Tv_{i,k},
	\end{equation}
	\begin{equation}
	\label{eq3}
	{v}_{i,k+1}={v}_{i,k}+Tacc_{i,k},
	\end{equation}
	\begin{equation}
	\label{eq4}
	{acc}_{i,k+1}=(1-\frac{T}{\tau_{i}})acc_{i,k}+\frac{T}{\tau_{i}}a^{\rm CL}_{i,k},
	\end{equation}
	\noindent where $\tau_{i}$ is a time constant representing driveline dynamics. The first-order-system approximation in \eqref{eq4} is widely used in platoon controller design, which is obtained by first formulating a non-linear model and then applying the exact feedback linearization technique to convert the non-linear model to a linear one \cite{zheng2015stability,Stankovic2000,Lin2021}.} In order to ensure driving safety and comfort, the following constraints are applied
	\begin{equation}
	\label{eq43}
	acc_{\mathrm{min}} \leq acc_{i,k}\leq acc_{\mathrm{max}},\ a^{\rm CL}_{\mathrm{min}} \leq a^{\rm CL}_{i,k}\leq a^{\rm CL}_{\mathrm{max}}
	\end{equation}
	\noindent where $acc_{\mathrm{min}}$ and $acc_{\mathrm{max}}$ are the acceleration limits, while $a^{\rm CL}_{\mathrm{min}}$ and 	$a^{\rm CL}_{\mathrm{max}}$ are the control input limits. \par
	
	The headway of follower $i$ at control interval $k$, i.e., bumper-to-bumper distance between follower $i$ and its predecessor $i-1$, is denoted by $d_{i,k}$ with
	\begin{equation}
	\label{eq5}
	d_{i,k}=p_{i-1,k}-p_{i,k}-L_{i-1},
	\end{equation}
	\noindent where $L_{i-1}$ is the the body length of vehicle $i-1$.
	
	We adopt the Constant Time-Headway Policy (CTHP), where follower $i$ aims to maintain the desired headway 
	\begin{equation}
	\label{eq6}
	d_{r,i,k}=r_{i}+h_{i}v_{i,k},
	\end{equation}
	\noindent where $r_{i}$ is a standstill distance for the safety of follower $i$ and $h_{i}$ is a constant time gap of follower $i$, which represents the time that it takes for follower $i$ to bridge the distance in between the vehicles $i$ and $i-1$ when continuing to drive with a constant velocity.
	
	The tracking errors, i.e., gap-keeping error $e_{pi,k}$ and velocity error $e_{vi,k}$ of follower $i$ are defined as
	\begin{equation}
	\label{Tra_err}
	e_{pi,k}=d_{i,k}-d_{r,i,k},\ e_{vi,k}=v_{i-1,k}-v_{i,k}.
	\end{equation}

\subsection{C-V2X Communications Module}
    \begin{figure}[tb!]
    \centering
    \includegraphics[width=0.48\textwidth]{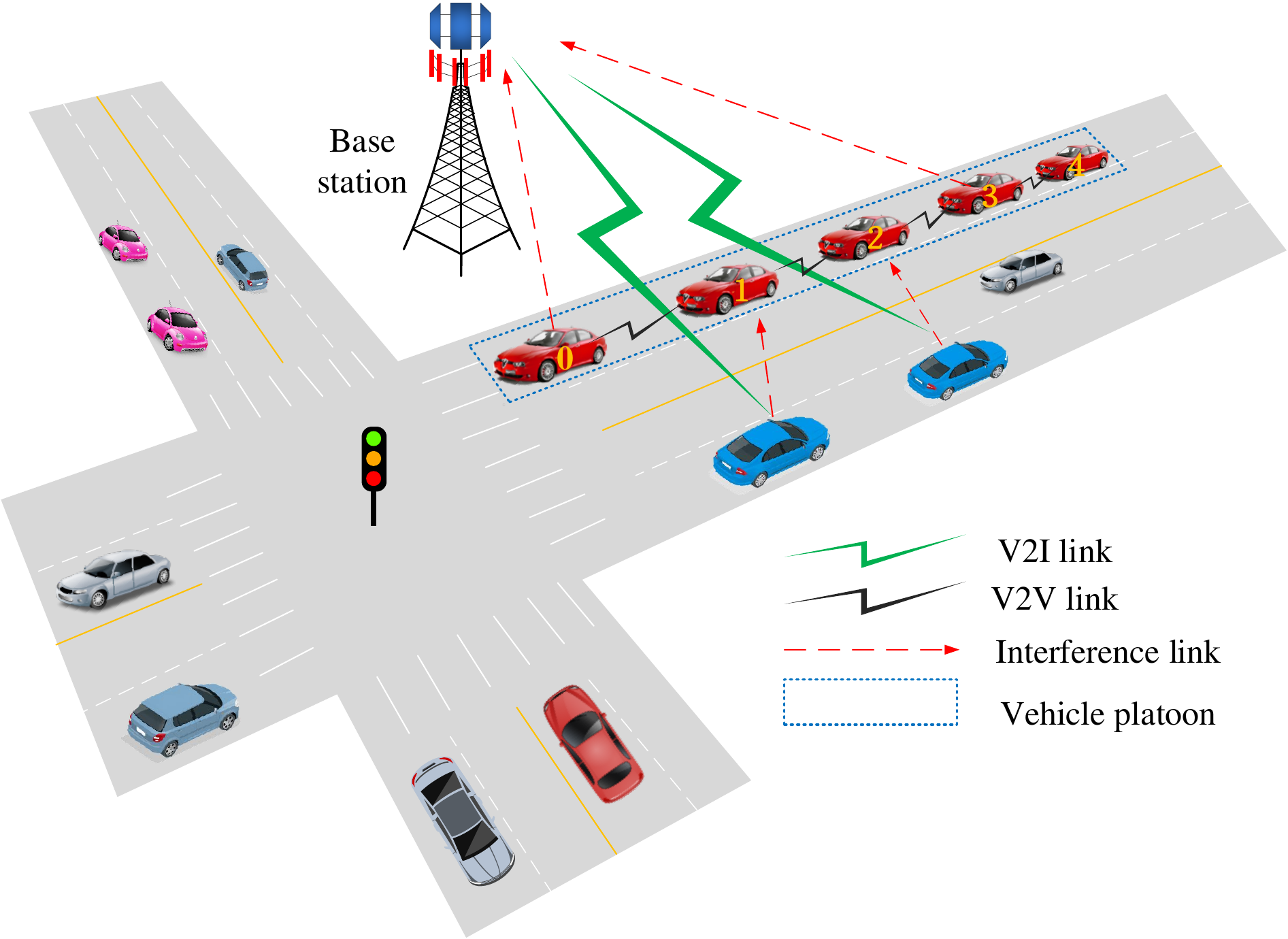}
    \caption{An illustration of C-V2X network for the urban environment}
    \label{fig_1}
    \end{figure}
As shown in Fig.~\ref{fig_1}, we consider a typical urban C-V2X network, where V2V links coexist with V2I links. A V2I link connects a vehicle to the Base Station (BS) and is used for high-throughput services. According to the PF IFT, a V2V link connects a pair of predecessor and follower for periodic transmission of CAM. The link between vehicle $i$ and $i+1$ is denoted as V2V link $i$. The set of $N-1$ V2V links can thus be represented by $\mathcal{V}\backslash\{N-1\}$. \par 

We consider that there are $M$ V2I links (uplink considered). Without loss of generality, we assume that every V2I link $m\in \mathcal{M}=\{0,\cdots, M-1\}$ is pre-assigned sub-channel $m$ with constant transmit power $P^{\rm I}_{m}$ \cite{liang2019spectrum}. In order to enhance spectrum utilization, {one or more V2V links} can reuse the sub-channels of the V2I links for CAM transmission. We use the binary allocation indicator $\theta_{i,m,(k,t)}\in\{0,1\}$ to indicate whether V2V link $i$ occupies sub-channel $m$ at communication interval $(k, t)$ or not. Moreover, we consider that each V2V link $i$ occupies at most one sub-channel, i.e., $\sum_{m=0}^{M-1} \theta_{i,m,(k,t)} \leq 1$.\par
At the beginning of each control interval $k$, each vehicle $i$ samples its driving status to form the CAM and buffers the CAM in a queue before transmitting the data to the following vehicle $i+1$. In each communication interval $(k,t)$, each vehicle $i$ transmits the data in its queue according to the local RRA decisions. 

\subsubsection{Channel gain}

The instantaneous channel gain of V2V link $i$ over sub-channel $m$ (occupied by V2I link $m$) at communication interval $(k,t)$ is denoted by $G_{i,m,(k,t)}$. Similarly, let $G_{m,(k,t)}$ denote the channel gain of the V2I link $m$; $G_{i,B,m,(k,t)}$ the interference channel gain from V2V link $i$ transmitter to V2I link $m$ receiver; $G_{B,i,m,(k,t)}$ the interference channel gain from V2I link $m$ transmitter to V2V link $i$ receiver; and $G_{j,i,m,(k,t)}$ the interference channel gain from the V2V link $j$ transmitter to the V2V link $i$ receiver over the sub-channel $m$.\par

\subsubsection{Signal-to-interference-plus-noise ratio (SINR)}
The SINR $\gamma_{m,(k,t)}$ of V2I link $m$ and the SINR $\gamma_{i,m,(k,t)}$ of V2V link $i$ on sub-channel $m$ at communication interval $(k,t)$ are derived by 
\begin{align}\label{SINRV2I}
\gamma_{m,(k,t)} = \frac{P^{\rm I}_{m} G_{m,(k,t)}}{\sigma^2 + \sum\limits_{i\in\mathcal{V}\backslash\{N-1\}}\theta_{i,m,(k,t)} P^{\rm V}_{i,m,(k,t)} G_{i,B,m,(k,t)}},
\end{align}
and
\begin{align}\label{SINRV2V}
\gamma_{i,m,(k,t)}= \frac{P^{\rm V}_{i,m,(k,t)} G_{i,m,(k,t)}}{\sigma^2 + I_{i,m,(k,t)}},
\end{align}
respectively,
\noindent where $P^{\rm V}_{i,m,(k,t)}$ is the transmit power of V2V link $i$ over the sub-channel $m$ at communication interval $(k,t)$. $\sigma^2$ is the power of channel noise which satisfies the independent Gaussian distribution with a zero mean value. $I_{i,m,(k,t)}$ is the total interference power received by V2V link $i$ over sub-channel $m$, where
\begin{align}\label{IV2V}
& I_{i,m,(k,t)}=\IEEEnonumber \\
& P^{\rm I}_{m} G_{B,i,m,(k,t)}+ \sum\limits_{j\in \mathcal{V} \backslash \{i,N-1\}}\theta_{j,m,(k,t)}  P^{\rm V}_{j,m,(k,t)} G_{j,i,m,(k,t)}.\IEEEnonumber
\end{align}
\subsubsection{Instantaneous data rate}

The instantaneous data rates $r_{m,(k,t)}$ and $r_{i,(k,t)}$ of V2I link $m$ and V2V link $i$ at communication interval $(k,t)$ are respectively derived as
\begin{equation}
    r_{m,(k,t)} = W\log_2(1+\gamma_{m,(k,t)}),
\end{equation}
and
\begin{align}\label{rateV2V}
r_{i,(k,t)} =\sum_{m=0}^{M-1} {\theta_{i,m,(k,t)}} W\log_2(1+\gamma_{i,m,(k,t)}),
\end{align}
\noindent where $W$ is the bandwidth of a sub-channel.\par
Let $r^{\rm CAM}_{i,(k,t)}$ denote the transmission rate of V2V link $i$ in terms of CAM at communication interval $(k,t)$, which is given by
\begin{equation}
	\label{rateV2VCAM}
        r^{\rm CAM}_{i,(k,t)}=\frac{r_{i,(k,t)}}{N_c},
\end{equation}
\noindent where $N_c$ is the constant CAM size. \par

\subsubsection{Queuing dynamic}
Each vehicle $i\in\mathcal{V}\backslash \{N-1\}$ except for the last vehicle $N-1$ has a buffer to store its CAM, where the buffer capacity is $N_Q$ in the number of CAM. Let $q^{\rm CAM}_{i,(k,t)}$ denote the queue length of vehicle $i$ in the number of CAM at communication interval $(k,t)$. If the queue length $q^{\rm CAM}_{i,(k,t)}$ reaches the buffer capacity $N_Q$, the subsequent arriving data will be dropped. The queue process evolves as
\begin{align}
    \label{queue}
    \setlength{\arraycolsep}{1.6pt}
        q^{\rm CAM}_{i,(k,t+1)}=
         \left\{
        \begin{array}{ll}
	\min \left[N_Q, \max [0,q^{\rm CAM}_{i,(k,t)}- 10^{-3}\times \right. \\ \left.r^{\rm CAM}_{i,(k,t)}]+1 \right], &\mathrm{if}\quad t=0 \\
	\max \left[0,q^{\rm CAM}_{i,(k,t)}- 10^{-3}\times r^{\rm CAM}_{i,(k,t)}\right], & \mathrm{otherwise} \\
	\end{array}\right. .
\end{align}
At each communication interval $(k, t)$, the queue length $q^{\rm CAM}_{i,(k,t+1)}$ is decreased by $10^{-3}\times r^{\rm CAM}_{i,(k,t)}$, which is the number of CAM transmitted during the communication interval. Meanwhile, at every communication interval $(k, 0)$, the queue length $q^{\rm CAM}_{i,(k,t+1)}$ is increased by $1$, since vehicle $i$ samples the driving status for control interval $k$ and buffers the generated CAM. In addition, the CAM that is not fully transmitted during control interval $k$ will continue to be transmitted in the next control interval $k+1$.  \par

\subsection{Correlation between Platoon Control Decisions and Radio Resource Allocation Decisions}

In our system model, each vehicle $i\in\mathcal{V}\backslash\{0\}$ makes PC decisions on the control input $a^{\rm CL}_{i,k}$ at every control interval $k\in\mathcal{K}$. Moreover, each vehicle $i\in\mathcal{V}\backslash\{N-1\}$ makes RRA decisions on sub-channel allocation $\{\theta_{i,m,(k,t)}\}_{m\in\mathcal{M}}$ and transmit power $\{P^{\rm V}_{i,m,(k,t)}\}_{m\in\mathcal{M}}$ at every communication interval $(k,t)$, where $k\in\mathcal{K}$ and $t\in\mathcal{T}$. It is important to note that the PC and RRA decisions are closely related to each other.\par

At the beginning of each control interval $k$, each follower $i\in\mathcal{V}\backslash\{0\}$ determines $a^{\rm CL}_{i,k}$ based on its own driving status as well as the driving status received from its predecessor $i-1$. Let $\tau_{i,k}$ be the observation delay of follower $i$ at control interval $k$. Thus, $c_{i-1,k-\tau_{i,k}}=\{p_{i-1,k-\tau_{i,k}},v_{i-1,k-\tau_{i,k}},acc_{i-1,k-\tau_{i,k}}\}$ is the most recent available delayed CAM at follower $i$, which correspond to the position, velocity, and acceleration sampled at predecessor $i-1$ in control interval $k-\tau_{i,k}$. Therefore, the observed driving status of vehicle $i$ is defined as
\begin{align}
x_{i,k-\tau_{i,k}}=&\{e_{pi,k-\tau_{i,k}},e_{vi,k-\tau_{i,k}},acc_{i,k-\tau_{i,k}},acc_{i-1,k-\tau_{i,k}} \},
\end{align}  
\noindent where $e_{pi,k-\tau_{i,k}}=p_{i-1,k-\tau_{i,k}}-p_{i,k-\tau_{i,k}}-L_{i-1}-d_{r, i,k}$, $e_{vi,k-\tau_{i,k}}=v_{i-1,k-\tau_{i,k}}-v_{i,k-\tau_{i,k}}$. Note that although follower $i$ has the undelayed observation on its own $p_{i,k}$, $v_{i,k}$ and $acc_{i,k}$, the observation $x_{i,k-\tau_{i,k}}$ is defined based on $p_{i,k-\tau_{i,k}}$, $v_{i,k-\tau_{i,k}}$ and $acc_{i,k-\tau_{i,k}}$ to be aligned with the delayed information from its predecessor $i-1$.  \par 

The observation delay $\tau_{i,k}$ depends on the transmission delay of CAM over V2V link $i-1$, which can be derived from $q^{\rm CAM}_{i-1,(k-1,T)}$ or $q^{\rm CAM}_{i-1,(k,0)}$ as
\begin{equation}
\label{eq23}
\tau_{i,k}=\lceil q^{\rm CAM}_{i-1,(k-1,T)} \rceil+1=\lceil q^{\rm CAM}_{i-1,(k,0)} \rceil+1.
\end{equation}  

\begin{figure}[tb!]
\centering

\includegraphics[width=0.48\textwidth]{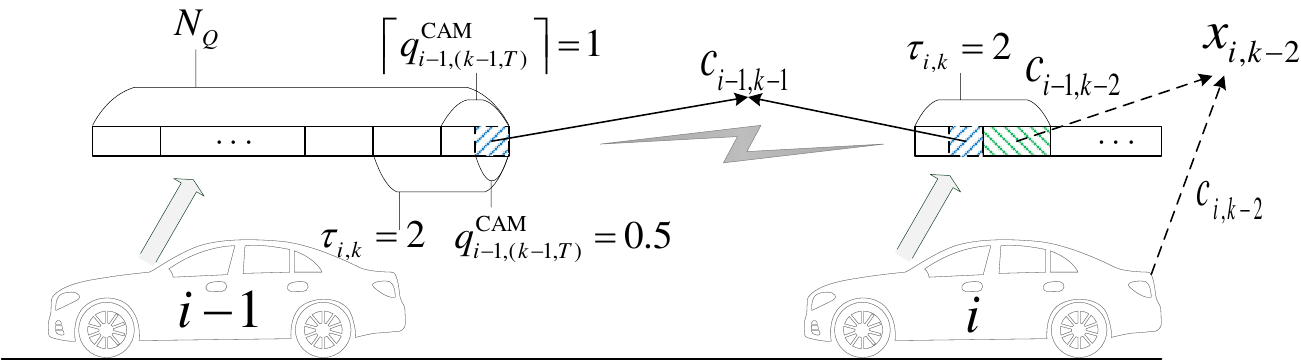}
\caption{A schematic diagram of the relationship between observation delay $\tau_{i,k}$ and queue length $q^{\rm CAM}_{i-1,(k-1,T)}$}
\label{fig_relationship}
\end{figure}

{An example is given in Fig.\ref{fig_relationship} to illustrate the relationship between the queue length and observation delay. Assume the queue length of predecessor $i-1$ at the end of control interval $k-1$ is $0.5$ CAM of $c_{i,k-1}$, i.e., $q^{\rm CAM}_{i-1,(k-1,T)}=0.5$. This means that the CAM $c_{i,k-1}$ generated by the predecessor $i-1$ at control interval $k-1$ is not fully received by the follower $i$. At control interval $k$, since the follower $i$ cannot interpret an incomplete CAM of $c_{i,k-1}$, it has to make decisions based on the last fully received CAM, i.e., the CAM $c_{i,k-2}$ generated by predecessor $i-1$ at control interval $k-2$. Therefore, the observation delay is $\tau_{i,k}=\lceil 0.5 \rceil+1=2$. The delayed observation for follower $i$ at control interval $k$ is $x_{i,k-2}$.}\par

{Please note that \eqref{eq23} no longer holds when the queue length $q^{\rm CAM}_{i,(k,t)}$ reaches the buffer capacity $N_Q$ and the subsequent arriving data are dropped. In this paper, we consider the case when $N_Q$ is large enough and the packet dropping probability is negligible. We leave the consideration of dropped packets to future work.}\par

{In this paper, we consider all the generated CAMs are buffered and transmitted sequentially. Another popular buffer management strategy is to replace any
old CAM that has not yet been fully delivered in the previous control interval with the newly generated CAM at the beginning of each control interval. We adopt the current strategy since the probability of successfully transmitting a partially transmitted CAM is larger than that of transmitting a completely new CAM due to the smaller amount of data left to be transmitted. However, our proposed MTCC framework can be applied with other buffer managements strategies with the change of \eqref{queue} and \eqref{eq23}}.\par

The correlation between PC decisions and RRA decisions can be analyzed from the following two aspects:	
\begin{description}
	\item[Impact of RRA on PC] The PC decisions are made with the target of optimizing the PC performance, which is affected by the observation delay $\tau_{i,k}$. Meanwhile, $\tau_{i,k}$ is determined by $q^{\rm CAM}_{i-1,(k-1,T)}$ according to \eqref{eq23}, which in turn depends on the RRA decisions  $\{\theta_{i,m,(k',t)}\}_{k'<k,m\in\mathcal{M}}$ and $\{P^{\rm V}_{i,m,(k',t)}\}_{k'<k,m\in\mathcal{M}}$ according to \eqref{SINRV2I}-\eqref{queue}. Different RRA decisions lead to diverse stationary distributions of observation delay. \emph{Therefore, the PC decisions should be optimized under the stochastic delay distributions stemmed from the de facto RRA decisions.}  
	\item [Impact of PC on RRA] The RRA decisions are made with the targets of (1) maximizing the V2I throughput; (2) minimizing the PC performance degradation due to delayed observation. Unfortunately, these two targets are contradictory with each other and an optimal trade-off should be struck. The trade-off heavily depends on the impact of observation delay $\tau_{i,k}$ on PC performance, which in turn is affected by the PC decisions. Better PC decisions lead to higher tolerance to observation delay, which means that larger V2I throughput can be supported with negligible penalty to PC performance. \emph{Therefore, the RRA decisions should be optimized with awareness of the impact of observation delay on PC performance under the de facto PC decisions.   }
\end{description}


\section{Communication-Aware DRL-based Platoon Control}
We assume that the RRA policy $\pi^{\rm CM}$ is available and focus on learning the PC policy $\pi^{\rm CL}_{i}, i\in\mathcal{V}\backslash\{0\}$. The PC problem with observation delay is essentially a Random Delay Decentralized Partially Observable Markov Decision Process (RD-Dec-POMDP). Each follower $i\in\mathcal{V}\backslash\{0\}$ is a PC agent, which makes a local and delayed observation at each control interval $k$, and decides on its local actions to maximize its expected cumulative \emph{individual} reward. The cumulative reward is normally referred to as the return in the RL literature.\par 

\subsection{PC State}	

The state for each PC agent $i$ at control interval $k$ is defined as 
\begin{equation}
\label{pc_state}
  S^{\rm CL}_{i,k}=\{x_{i,k-\tau_{i,k}},\{a^{\rm CL}_{i,k'}\}_{k'=k-\tau_{\mathrm{max}}}^{k-1},\tau_{i,k}\},  
\end{equation}
\noindent {where $\tau_{\mathrm{max}}$ is the maximum observation delay that depends on the maximum queue length $N_{Q}$, i.e., $\tau_{\mathrm{max}} \geq \tau_{i,k}$}. The delayed observation of driving status $x_{i,k-\tau_{i,k}}$ is augmented with the last $\tau_{\mathrm{max}}$ actions $\{a^{\rm CL}_{i,k'}\}_{k'=k-\tau_{\mathrm{max}}}^{k-1}$ of PC agent $i$. Moreover, the observation delay $\tau_{i,k}$ is included since it provides useful information to the control agent on how old an observation is. More importantly, $\tau_{i,k}$ serves as a bridge between the control and communication modules. For PC agent $i$ to be aware of the observation delay $\tau_{i,k}$, the predecessor $i-1$ needs to share its queue length $q_{i-1,(k,0)}^{\mathrm{CAM}}$ at the beginning of each control interval $k$ via control signaling. This is possible in the C-V2X system, since the queue length can be contained in Sidelink Control Information (SCI) transmitted in the Physical Sidelink Shared Channel (PSSCH). \par

\subsection{PC Action}	
The control input, $a^{\rm CL}_{i,k} \in [a^{\rm CL}_{\mathrm{min}},a^{\rm CL}_{\mathrm{max}}]$ of each PC agent $i$ is regarded as its PC action at control interval $k$. 

\subsection{PC Reward Function}	
The objective for each PC agent $i$ is to minimize its own gap-keeping error $e_{pi,k}$ and velocity error $e_{vi,k}$ while penalizing control input $a^{\rm CL}_{i,k}$ and the jerk to reduce the fuel consumption and improve the driving comfort, respectively. Note that the jerk is the change rate in acceleration, which is given by 
\begin {align}	
	j_{i,k}&=\frac {acc_{i,k+1}-acc_{i,k}}{T}=-\frac{1}{\tau_{i}}acc_{i,k}+\frac{1}{\tau_{i}}a^{\rm CL}_{i,k}, 
\end{align}
\noindent where the second equality is due to the forward Euler discretization of \eqref{eq4}.\par

The individual reward for each PC agent $i$ is given by 
\begin{align}
\label{PC_Reward}
&	R^{\rm CL}_{i,k}(x_{i,k},a^{\rm CL}_{i,k})= \IEEEnonumber \\
&-\{|\frac{e_{pi,k}}{\hat{e}_{p,\mathrm{max}}}|+ {\alpha_1}|\frac{e_{vi,k}}{\hat{e}_{v,\mathrm{max}}}|+ {\alpha_2}|\frac{a^{\rm CL}_{i,k}}{a^{\rm CL}_{\mathrm{max}}}|+ {\alpha_3}|\frac{j_{i,k}}{2acc_{\mathrm{max}}/T}|\},
\end{align}
 where $\hat{e}_{p,\mathrm{max}}$ and $\hat{e}_{v,\mathrm{max}}$ are the nominal maximum control errors such that it is larger than most possible control errors.  {$\alpha_1$},  {$\alpha_2$}, and  {$\alpha_3$} are the positive weights and can be adjusted to determine the relative importance of minimizing the gap-keeping error, the velocity error, the control input, and the jerk. \par

The expected return $J^{\rm CL}_{i}$ of PC agent $i$ under policy $\pi^{\rm CL}_{i}$ can be expressed as 
\begin{equation}
\label{pc_perf}
		J^{\rm CL}_{i}=\mathrm{E}_{\pi^{\rm CM}}\mathrm{E}_{\pi^{\rm CL}_{i}}[\sum_{k=0}^{K-1} \gamma^{k} R^{\rm CL}_{i,k}],\ 0 \leq \gamma \leq 1,
\end{equation}
\noindent where $\gamma$ is the PC reward discount factor and $\pi^{\rm CM}$ is the de facto communication policy. 

\newtheorem{remark}{Remark}
\begin{remark}[Impact of communications policy on control performance] In \eqref{pc_perf}, the expectation is taken with respect to the probability distribution of the state-action trajectories when the PC agent $i$ follows policy $\pi_{i}^{\mathrm{CL}}$ and the RRA policy is $\pi^{\rm CM}$. The RRA policy $\pi^{\rm CM}$ affects the PC performance since it determines the observation delay $\tau_{i,k}$, which is a part of the augmented state $S^{\rm CL}_{i,k}$. In other words, $\pi^{\rm CM}$ has an important influence on the state transition probabilities of the RD-Dec-POMDP model.
\end{remark}

The objective of the PC problem is for each PC agent $i$ to find the optimal policy $\pi^{\rm CL*}_{i}$ under delayed observation that maximizes its individual expected return $J^{\rm CL}_{i}$, i.e.,
\begin{equation}
		\label{eq16}
		\pi^{\rm CL*}_{i}=\arg\max_{\pi^{\rm CL*}_{i}}J^{\rm CL}_{i}, \ \forall i\in\mathcal{V}\backslash\{0\}.
\end{equation}

\section{DRL Solution}	

The DDPG algorithm \cite{lillicrap2015continuous} is utilized to solve the PC problem, which is the most extensively used algorithm in the existing DRL-based car-following controllers. Since DDPG is designed to solve MDP problems, it is questionable whether the algorithm is suitable for solving the RD-Dec-POMDP problem of PC. In the following, we discuss the adoption of DDPG in the multi-agent setting and random delay setting, respectively. \par

\subsection{Multi-agent problem in DRL-based PC}
The PC problem corresponds to a Dec-POMDP and lies in the multi-agent domain. Although there are various multi-agent algorithms such as MADDPG \cite{lowe2017multi} for applying RL to multi-agent systems, we adopt the IL approach where each agent learns independently using DDPG. The reason for choosing IL is due to its simplicity and scalability. More importantly, the non-stationary environment issue for IL is greatly alleviated in the PC problem, since it is proved in \cite{lei2022deep} that only the actions of its predecessors but not the followers will affect the environment of a PC agent. Furthermore, the credit assignment issue in multi-agent problem does not exist for our PC model, as each agent optimizes its individual return instead of the global return that is the sum of individual returns over all the PC agents.  


\subsection{Random observation delay problem in DRL-based PC}
The theoretical foundation of DDPG algorithm is the Deterministic Policy Gradient (DPG) Theorem \cite{lillicrap2015continuous,silver2014deterministic}, which shows that deterministic policy gradient is the expected gradient of the action-value function for any MDP whose corresponding gradients exist. By the discussion in Section V.A, we can approximately consider that the undelayed driving status $x_{i,k}$ at PC agent $i$ is Markov, i.e., $p(x_{i,k+1}|x_{i,k},a_{i,k})=p(x_{i,k+1}|\dots,x_{i,k-1},x_{i,k},a_{i,k})$, ignoring the impact of the predecessors' actions on $x_{i,k+1}$. However, each PC agent $i$ can only observe the delayed driving status $x_{i,k-\tau_{i,k}}$ instead of $x_{i,k}$, where $x_{i,k-\tau_{i,k}}$ is no longer a Markov state. It is proved in the following Theorem 1 that $S_{i,k}^{\mathrm{CL}}$ becomes a Markov state by augmenting the delayed observation of driving status with action history.

\newtheorem{theorem}{Theorem}
	\begin{theorem}
		Markov property is ensured for the augmented state $S_{i,k}^{\mathrm{CL}}$, i.e., $p(S_{i,k+1}^{\mathrm{CL}}|S_{i,k}^{\mathrm{CL}},a^{\rm CL}_{i,k})=p(S_{i,k+1}^{\mathrm{CL}}|\dots,S_{i,k-1}^{\mathrm{CL}},S_{i,k}^{\mathrm{CL}},a^{\rm CL}_{i,k})$.
	\end{theorem}

The proof of Theorem 1 is given in Appendix A. \par

The reward function $R^{\rm CL}_{i,k}(x_{i,k},a^{\rm CL}_{i,k})$ defined in \eqref{PC_Reward} is a function of the undelayed observation $x_{i,k}$ instead of the augmented state $S_{i,k}^{\rm CL}$. In order to construct an MDP for the delayed observations, we define the delayed reward function $\tilde{R}^{\rm CL}_{i,k}(S_{i,k}^{\rm CL},a^{\rm CL}_{i,k})$ for each follower $i\in\mathcal{V}\backslash\{0\}$ as the expected reward obtained by PC agent $i$ in augmented state $S_{i,k}^{\rm CL}$, i.e.,
\begin{equation}
\tilde{R}^{\rm CL}_{i,k}(S_{i,k}^{\rm CL},a^{\rm CL}_{i,k})=\mathrm{E}_{x_{i,k}}[R^{\rm CL}_{i,k}(x_{i,k},a^{\rm CL}_{i,k})|S_{i,k}^{\rm CL}].
\end{equation}

Now we construct the augmented state MDP $\tilde{\mathcal{M}}_{i}=(S_{i,k}^{\rm CL},a^{\rm CL}_{i,k},\tilde{R}^{\rm CL}_{i,k},p,\gamma)$. Note that under $\tilde{\mathcal{M}}_{i}$, the expected return $\tilde{J}^{\rm CL}_{i}$ of PC agent $i$ under policy $\pi^{\rm CL}_{i}$ is written as
\begin{equation}
\label{pc_perf_aug}
\tilde{J}^{\rm CL}_{i}=\mathrm{E}_{\pi^{\rm CM}}\mathrm{E}_{\pi^{\rm CL}_{i}}[\sum_{k=0}^{K-1} \gamma^{k} \tilde{R}^{\rm CL}_{i,k}], 
\ 0 \leq \gamma \leq 1.
\end{equation}  

Thus, the optimal policy for $\tilde{\mathcal{M}}_{i}$ is given as
\begin{equation}
\label{optimal_policy_aug}
\tilde{\pi}^{\rm CL*}_{i}=\arg\max_{\pi^{\rm CL*}_{i}}\tilde{J}^{\rm CL}_{i}, \ \forall i\in\mathcal{V}\backslash\{0\}.
\end{equation}   

The following Theorem 2 states that the optimal policy $\tilde{\pi}^{\rm CL*}_{i}$ for the augmented state MDP $\tilde{\mathcal{M}}_{i}$ is the same as the optimal policy $\pi^{\rm CL*}_{i}$ in \eqref{eq16} for our PC problem under delayed observation.\par

\newtheorem{theorem2}[theorem]{Theorem}
\begin{theorem2}
If the initial distributions of $p(x_{i,0})$ and $p(S_{i,0}^{\rm CL})$ satisfy 
\begin{equation}
\label{distribution_x}
p(x_{i,0})=p(S_{i,0}^{\rm CL})\mathrm{E}_{\pi^{\rm CM}}\mathrm{E}_{\pi^{\rm CL}_{i}}[\mathbf{1}(x_{i,0})|S_{i,0}^{\rm CL}],
\end{equation}
\noindent we have
\begin{equation}
\tilde{J}^{\rm CL}_{i}=J^{\rm CL}_{i}, \ \tilde{\pi}^{\rm CL*}_{i}=\pi^{\rm CL*}_{i}
\end{equation}
\end{theorem2}

The proof of Theorem 2 is given in Appendix B.\par
 
Based on Theorem 2, the optimal PC policy under delayed observation $\pi^{\rm CL*}_{i}$ can be derived by solving $\tilde{\mathcal{M}}_{i}$. For this purpose, we apply the DDPG algorithm and the deterministic policy gradient for $\tilde{\mathcal{M}}_{i}$ is given in Lemma 1.
\newtheorem{lemma}{Lemma}
\begin{lemma}
		The deterministic policy gradient for the augmented state MDP $\tilde{\mathcal{M}}_{i}$ is
		\begin{align}
		\label{DPG}
		\bigtriangledown_{\theta^{\mu}_{i}}J_{i}^{\rm CL}(\mu^{\rm CL}_{{i}})&=\mathrm{E}[\bigtriangledown_{\theta^{\mu}_{i}}\mu^{\rm CL}_{i}(S^{\rm CL}_{i,k}|\theta^{\mu}_{i}) \IEEEnonumber \\
	&\bigtriangledown_{a}Q^{\rm CL}_{i}(S^{\rm CL}_{i,k}, a |\theta^{Q}_{i})|_{a=\mu^{\rm CL}_{i}(S^{\rm CL}_{i,k}|\theta^{\mu}_{i}) }].
		\end{align}
\end{lemma}
The proof of Lemma 1 is straightforward as $\tilde{\mathcal{M}}_{i}$ is an MDP for which the DPG Theorem can be directly applied. \par

In order to sample the deterministic policy gradient in \eqref{DPG}, we need to evaluate the action-value function $Q_{\mu_{\theta_{i}}}^{\rm CL}(S^{\rm CL}_{i,k},a^{\rm CL}_{i,k})$ of the augmented state MDP $\tilde{\mathcal{M}}_{i}$. Based on the following Bellman equation 
\begin{align}
\label{augmentQ}
&Q_{\mu_{\theta_{i}}}^{\rm CL}(S^{\rm CL}_{i,k},a^{\rm CL}_{i,k})=\mathrm{E}_{x_{i,k}}[R^{\rm CL}_{i,k}(x_{i,k},a^{\rm CL}_{i,k})|S_{i,k}^{\rm CL}]  \IEEEnonumber \\
&+\gamma \mathrm{E}_{S_{i,k+1}^{\mathrm{CL}}}[Q_{\mu_{\theta_{i}}}^{\rm CL}(S^{\rm CL}_{i,k+1},\mu_{\theta_{i}}(S^{\rm CL}_{i,k+1}))|S_{i,k}^{\mathrm{CL}},a^{\rm CL}_{i,k}],
\end{align}
\noindent the PC agent $i$ can sample the undelayed reward $R^{\rm CL}_{i,k}(x_{i,k},a^{\rm CL}_{i,k})$ and next state $S^{\rm CL}_{i,k+1}$ at control interval $k$, and calculate the temporal-difference (TD) target as
\begin{equation}
\label{TD}
y_{i,k}=R^{\rm CL}_{i,k}(x_{i,k},a^{\rm CL}_{i,k})+\gamma Q_{\mu_{\theta_{i}}}^{\rm CL}(S^{\rm CL}_{i,k+1},\mu_{\theta_{i}}(S^{\rm CL}_{i,k+1})).
\end{equation}

\newtheorem{remark2}[remark]{Remark}
\begin{remark2}[Assumption of undelayed reward]
	We assume that there is no reward delay, i.e., the reward $R^{\rm CL}_{i,k}(x_{i,k},a^{\rm CL}_{i,k})$ based on the current driving status $x_{i,k}$ is available to the PC agent during training at each control interval $k$. This is possible since learning can be performed in a simulator or a laboratory in which the undelayed reward is available. After the agent learns the PC policy, the reward is no longer needed during execution when the undelayed reward is not available. 
\end{remark2}

\subsection{MTCC-PC Algorithm}
Based on the above discussion, the MTCC-PC algorithm is proposed. Each PC agent $i$ adopts the DDPG algorithm given in \cite{lillicrap2015continuous}. Specifically, DDPG develops both a pair of actor and critic networks, i.e., $\mu^{\rm CL}_{i}(S^{\rm CL}_{i,k}|\theta^{\mu}_{i})$ and $Q^{\rm CL}_{i}(S^{\rm CL}_{i,k}, a^{\rm CL}_{i,k} |\theta^{Q}_{i})$, to derive the optimal policy $\mu^{\rm CL*}_{i}(S^{\rm CL}_{i,k}|\theta^{\mu}_{i})$ and the corresponding action-value $Q^{\rm CL*}_{i}(S^{\rm CL}_{i,k}, a^{\rm CL}_{i,k} |\theta^{Q}_{i})$, respectively. A copy of the actor and critic networks are created as target networks, i.e., $\mu^{\rm CL'}_{i}(S^{\rm CL}_{i,k}|\theta^{\mu'}_{i})$ and $Q^{\rm CL'}_{i}(S^{\rm CL}_{i,k}, a^{\rm CL}_{i,k} |\theta^{Q'}_{i})$, to calculate the target values. To enable stable and robust learning, DDPG uses experience replay, and the networks are updated using minibatch samples from the experience buffer. During training, the sampled deterministic policy gradient ascent on $Q^{\rm CL}_{i}(S_{i,k},\mu^{\rm CL}_{i}(S_{i,k}|\theta^{\mu}_{i})|\theta^{Q}_{i})$ with regard to $\theta^\mu_{i}$ is used to train the actor network, and the critic network is trained by minimizing the Root Mean Square Error (RMSE) $L_{i,k} = y_{i,k}-Q^{\rm CL}_{i}(S^{\rm CL}_{i,k}, a^{\rm CL}_{i,k} |\theta^{Q}_{i})$ using the sampled gradient descent with respect to $\theta^{Q}_{i}$. { We refer the interested readers to \cite{lillicrap2015continuous} for the details of the DDPG algorithm.}  \par


{ In the following Remark 3 and Remark 4, we highlight two important design details of the MTCC-PC algorithm.}
\newtheorem{remark3}[remark]{Remark}
\begin{remark3}[Finite-horizon problem in DRL-based PC]
The PC problem in Section III.A considers a finite horizon with $K$ control intervals. However, the optimal policies are normally time-dependent in a finite-horizon setting, while DDPG is designed to solve the infinite-horizon or indefinite-horizon problems, where the learned policy is the same for every time step \cite{lei2020deep}. In order to deal with this problem, we set the target values of DDPG in the last control interval $K-1$ to be derived by \eqref{TD} in the way as for the other control intervals, i.e., the sum of the immediate reward and the discounted target Q value of the next state instead of only the immediate reward $R^{\rm CL}_{i,K-1}(x_{i,K-1},a^{\rm CL}_{i,K-1})$. Thus, the PC problem is transformed from a finite horizon problem to an infinite horizon problem. 
\end{remark3}

\newtheorem{remark4}[remark]{Remark}
\begin{remark4}[Simulation of delayed environment when training MTCC-PC]
The proposed MTCC-PC algorithm is is trained in a delayed environment generated by the simulation of C-V2X communications with de facto RRA policy rather than by a coarse-grained stochastic delay model. This is to ensure the delay distribution in the training environment is the same as that in the execution environment in practice. \par
\end{remark4}


\section{Experimental Results}
In this section, we design experiments to demonstrate that the proposed MTCC-PC algorithm outperforms the state-of-the-art communication-aware control, where the former is trained by a delayed environment generated by the fine-grained embedded simulation of C-V2X communications while the latter is trained by a simple stochastic delay model. Specifically, the baseline algorithm is Random Delay-aware PC (RD-PC), where the observation delay when training DRL-based PC is assumed to follow uniform distribution within the delay set $\{1,2,3,4,5\}$. { In addition, to demonstrate that MTCC-PC can improve the PC performance by augmenting the PC state with action history, we design a baseline algorithm, namely PC without augmented state (PC\_wo\_AS), which is the same as MTCC-PC except that the state only includes the delayed observation of driving state $x_{i,k-\tau_{i,k}}$.} MTCC-PC, RD-PC, and PC\_wo\_AS are both trained for $10000$ episodes and tested where C-V2X communications are implemented with the random RRA policy. Therefore, the induced delay distribution is the same for RD-PC, PC\_wo\_AS, and MTCC-PC when evaluating their performance.

\subsection{Experimental Setup}
\subsubsection{Driving data for leading vehicle 0}
All the DRL algorithms are trained/tested where the velocity profile of leading vehicle $0$ is obtained from the open-source driving data in \cite{Meixin2020}. Specifically, the driving data from the Next Generation Simulation (NGSIM) dataset \cite{NGSIM2009} was first obtained, based on which the car-following events were extracted by applying a car-following filter as described in \cite{Wang2018}. In our experiments, the velocity of the leading vehicle $0$ in each control episode follows the corresponding data of the leading vehicle in one car-following event, so that the real-world PC environment with uncertainty can be simulated. We used $900$ car-following events, $800$ of which are used for training and $100$ for testing. 
 
\subsubsection{Parameter setting}
The technical constraints and operational parameters of the PC and RRA environment are given in Table~\ref{table_1}. In general, the parameters of the PC environment are determined mainly using the values reported in \cite{Lin2021} and the urban case defined in \cite{3GPPr14v2x}. Each control episode is comprised of $120$ control intervals (i.e., $K=120$), where each control interval is set to $T = 0.05$ $\rm s$ \cite{wang2013self,buechel2018deep,ploeg2011design}. As the number of vehicles simulated in the existing literature on PC normally ranges from $3$ to $8$\cite{wang2018novel,Chu2019,Yan2021}, we set the number of vehicles to $N=5$. We initialize the driving status for the platoon with two-dimensional positions $\{p_{\rm V,i,0}\}_{0}^{N-1}=\{(416,427.5),(399,427.5),(383,427.5),(366, 427.5),(350,$ $427.5)\}$, $\{v_{i,0}\}^{N-1}_{i=0}=\{10,10,10,10,10\}$ $\rm m/s$, and $\{acc_{i,0}\}^{N-1}_{i=0}=\{0,0,0,0,0\}$ $\rm m/s^2$. Note that the two-dimensional positions $\{p_{\rm V,i,0}\}_{0}^{N-1}$ are used for RRA and the corresponding one-dimensional positions $\{p_i,0\}_{0}^{N-1}=\{416,399,383,366,350\}$ are used for 
 PC. For the V2I vehicles, we initialize them with two-dimensional positions $\{p_{\rm I,i,0}\}_{0}^{1}=\{(391, 434.75),(358, 434.75)\}$ and constant velocity $10$ $\rm m/s$. The nominal maximum control errors in the reward function \eqref{PC_Reward} are set to $\hat{e}_{p,\rm max}=10$ $\rm m$ and $\hat{e}_{v,\rm max}=10$ $\rm m/s$ so that it is larger than most possible control errors during training for all DRL algorithms. For the parameter setting of the RRA environment, we mainly follow the experimental setup in \cite{liang2019spectrum} for channel models of V2I and V2V links. The bandwidth of each sub-channel is set to $W=180$ $\rm kHz$.
\begin{table}[htb!]
        \renewcommand{\arraystretch}{1.1}
    \setlength\tabcolsep{1.6pt}  
    \centering
    \caption{Technical constraints and operational parameters of the PC and RRA environment}
    \begin{tabular}{ll}
        \hline
     \textbf{Description}&\textbf{Value} \\
        \hline
       \hline
        \multicolumn{2}{c} {\textbf{PC environment}}\\
        Control interval & $0.05$ $\rm s$\\
        Total time steps in each control episode $K$ &$120$\\
        Number of vehicles $N$&5\\
        Driveline dynamics time constant  $\tau_i$ & $0.1$ $\rm s$\\
        Time gap $h_i$ & $1$ $\rm s$\\
       Standstill distance $r_i$ & $2$ $\rm m$\\
        Body length of the vehicle $L_i$ &$4.5$ $\rm m$\\
        \hline
        Acceleration limitations $[acc_{\rm min},acc_{\rm max}]$& $[-4.3, 2.9]$  $\rm{m/s^2}$\\
        Control input limitations $[u_{\rm min},u_{\rm max}] $ & $[-4.3, 2.9]$  $\rm{m/s^2}$\\
        \hline
        Control reward coefficient $\{ {\alpha_1,\alpha_2,\alpha_3}\}$& $\{0.2, 0.1, 0.4\}$\\
        Nominal maximum gap-keeping error $\hat{e}_{p,\rm max}$ & $10$ $\rm{m}$\\
        Nominal maximum velocity error $\hat{e}_{v,\rm max}$ & $10$ $\rm{m/s}$\\
        \hline
       \hline
      \multicolumn{2}{c} {\textbf{RRA environment}}\\
        Communication interval &$1$ $\rm ms$\\
        Total time steps in each control interval $T$ &$50$\\
       Number of V2I links $M$ & $2$ \\
       Carrier frequency $f_c$& $2$ $\rm GHz$ \\
       Bandwidth of sub-channel $W$ & $180$ $\rm KHz$\\
       Noise power $\sigma^2$ &  $-114$ $\rm dBm $\\ 
       CAM size $N_c$ & $400$ bytes \\
       V2I transmit power $P^{\rm I}_{m}$& $23$ $\rm dBm$ \\
       V2V transmit power $P^{\rm V}_{i,m,(k,t)}$& $\{23,15,5,-100\}$ $\rm dBm$ \\ 
       BS antenna height & $25$ $\rm m$ \\
       BS antenna gain & $8$ $\rm dBi$ \\ 
       BS receiver noise figure & $5$ $\rm dB$ \\ 
       Vehicle antenna height & $1.5$ $\rm m$ \\
       Vehicle antenna gain & $3$ $\rm dBi$ \\
       Vehicle receiver noise figure & $9$ $\rm dB$ \\
       Communication reward coefficient $\{\kappa_1,\kappa_2\}$ &$\{0.001/W, 100\}$\\
       \hline
       \hline
    \end{tabular}
    \label{table_1}
\end{table}
	
The main hyper-parameters for training are summarized in Table \ref{hyper_parameters}. The values of all the hyper-parameters were selected by performing a grid search as in \cite{mnih2015human}, using the values reported in\cite{lillicrap2015continuous} as a reference. {RD-PC, PC\_wo\_AS and MTCC-PC algorithms have the same network architecture for DDPG, which has two hidden layers with $256$ and $128$ nodes, respectively.} The sizes of input layer is decided by the PF IFT. Moreover, an additional $1$-dimensional action input is fed to the second hidden layer for each critic network. The soft target update is implemented with a parameter of $0.001$.
\begin{table}[htbp!]
	\renewcommand{\arraystretch}{1.2}
	\caption{Hyper-Parameters of the DRL algorithms for training} \label{alg_para} \centering
	\begin{tabular}{ll}
		\hline
		{\textbf{Parameter}} & \textbf{Value} \\
		\hline
		Actor network size &$256,128$ \\
		\hline
		Critic network size & $256,128$ \\
		\hline
		Actor activation function & relu, relu, tanh \\ 
		\hline
		Critic activation function & relu, relu, linear\\ 
		\hline
		Actor learning rate $\psi$ & { $0.0001$}   \\
		\hline
		Critic learning rate $\varphi$ & { $0.001$} \\
		\hline
		Batch size $N_b$ & $64$  \\
		\hline
		Replay buffer size & $600000$  \\
		\hline
		Reward discount factor $\gamma$ &0.99 \\
		\hline
		Soft target update of DDPG &$0.001$ \\
        \hline
		Noise type &\makecell[l]{Ornstein-Uhlenbeck Process \\with $\theta=0.15$ and $\sigma=0.5$} \\
		\hline
		\makecell[l]{Final layer \\weights/biases initialization} & \makecell[l]{Random uniform distribution \\$[-3\times10^{-3},3\times10^{-3}]$} \\
		\hline
		\makecell[l]{Other layer \\weights/biases initialization} &\makecell[l]{Random uniform distribution$[-\frac{1}{\sqrt{f}},\frac{1}{\sqrt{f}}]$\\($f$ is the fan-in of the layer)} \\
		\hline
		\label{hyper_parameters}
	\end{tabular}
\end{table}

\subsection{Performance Comparison of MTCC-PC, RD-PC and PC\_wo\_AS}

\subsubsection{Performance for testing data}
The individual PC performance of each follower $ i \in\{1,2,3,4\}$ as well as the sum PC performance of the $4$ followers are reported in Table~\ref{table_PC_Performance_Step1} for MTCC-PC, RD-PC, and PC\_wo\_AS, respectively. The individual and the sum PC performance are obtained by averaging the returns of the corresponding followers and the sum returns of all followers, respectively, over $100$ test episodes after training is completed. Note that the return is the cumulative PC reward given in \eqref{PC_Reward} of one control episode. Compared with RD-PC and PC\_wo\_AS, MTCC-PC consistently shows better individual PC performance for each follower $i\in\{1,2,3,4\}$. Moreover, MTCC-PC outperforms RD-PC by $51.76\%$ in terms of the sum PC performance of all followers. It demonstrates that training in a delayed environment generated by embedded simulation of C-V2X communications rather than by a simple stochastic delay model can improve PC performance. { In addition, MTCC-PC outperforms PC\_wo\_AS by $46.46\%$ in terms of the sum PC performance of all followers, demonstrating that ensuring the Markov property by augmenting the PC state with action history can significantly improve the PC performance. }\par

\begin{table*}[htb!]
	\renewcommand{\arraystretch}{1}
	\setlength{\extrarowheight}{1pt}
	\centering
	\caption{PC performance after training with NGSIM dataset in one iteration. We present the individual performance of each follower as well as the sum PC performance of the $4$ followers for MTCC-PC and RD-PC, respectively.}
	\begin{tabular}{ |c|cccc|c|}
		\hline
		\multirow{2}{*}{\textbf{Algorithm}}&\multicolumn{4}{c|}{\textbf{ Individual PC performance}}&\multirow{2}{*}{\textbf{Sum PC performance}}\\
		\cline{2-5}
		&\textbf{Follower 1}&\textbf{Follower 2}&\textbf{Follower 3}&\textbf{Follower 4}&\\
		\hline
		MTCC-PC&-0.8029 &-0.5434 &-0.3214 &-0.3163 &-1.9840\\
		\cline{1-6}
		RD-PC&-1.4189&-1.1127 &-0.8573 &-0.7152 &-4.1041 \\
            \cline{1-6}
		{ PC\_wo\_AS}&{ -1.1750}&{ -0.9786} &{ -1.020} &{ -0.5318} &{ -3.7054} \\
		\hline	
	\end{tabular}
	\label{table_PC_Performance_Step1}
\end{table*}

\subsubsection{Convergence properties}
\begin{figure}[tb!]
\centering
\includegraphics[width=0.5\textwidth]{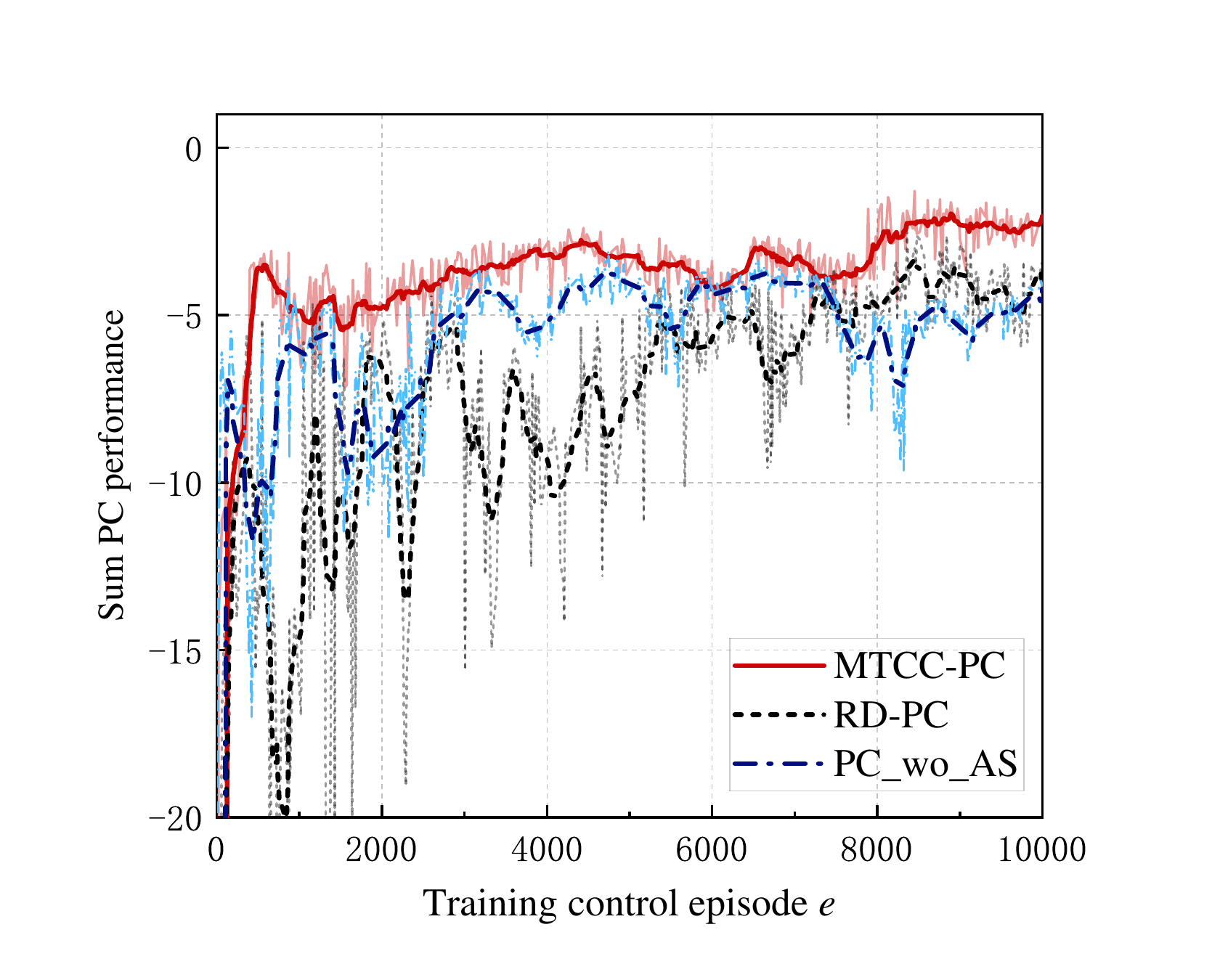}
\caption{Sum PC Performance during MTCC-PC, RD-PC, and PC\_wo\_AS training in one iteration. The vertical axis corresponds to the average returns over 10 test episodes. The dark curves correspond to smoothed curves and the light color curves correspond to the original curves.}
\label{fig_pc_convergence}
\end{figure}

The sum PC performance of MTCC-PC, RD-PC, and PC\_wo\_AS algorithms are evaluated periodically during training by testing in a delayed environment with fine-grained simulation of C-V2X communications under random RRA policy. Specifically, we run $10$ test episodes after every $10$ training episodes and average the sum PC performance over the $10$ test episodes as the performance for the latest $10$ training episodes. The performance as a function of the number of training episodes for MTCC-PC, RD-PC, and PC\_wo\_AS is plotted in Fig.~\ref{fig_pc_convergence}. { It can be observed from Fig.~\ref{fig_pc_convergence} that the performance of MTCC-PC is consistently better than those of RD-PC and PC\_wo\_AS during the whole training episode.} In addition, the performance curve of RD-PC exhibits significantly larger oscillation during all the training episodes compared to that of MTCC-PC, demonstrating that the convergence of RD-PC is relatively unstable. { Moreover, MTCC-PC has a significantly higher convergence rate than RD-PC and PC\_wo\_AS, as the performance of MTCC-PC converges at around $600$ episodes, while those of RD-PC and PC\_wo\_AS converge at around $8000$ and $3000$ episodes, respectively.} As explained above, MTCC-PC performs better than RD-PC since it is trained in an environment whose delay distribution is identical to that of the testing environment. Moreover, the faster and more stable convergence of MTCC-PC over RD-PC is also attributed to the fact that the observation delay in C-V2X communications is correlated between adjacent control intervals, while those generated by the uniform distribution are independent between control intervals. { In addition, MTCC-PC performs better than PC\_wo\_AS since it ensures the Markov property of the augmented PC state $S_{i,k}^{\mathrm{CL}}$.}

\begin{figure*}[tb!]
	\centering 
	\subfigure[MTCC-PC]{
			\includegraphics[scale=0.25]{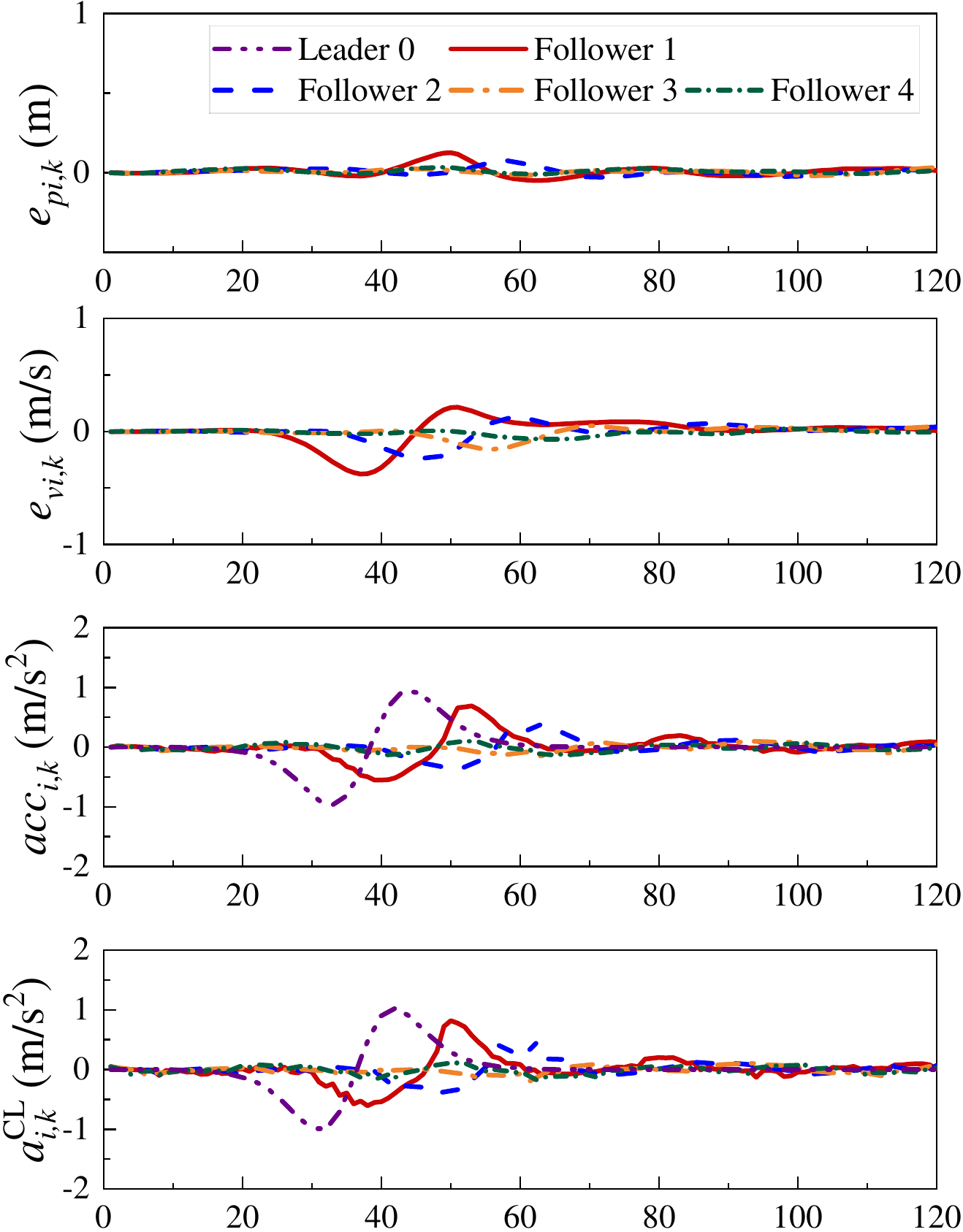}
		\label{MTCC_1}
	}	
	\subfigure[RD-PC]{
			\includegraphics[scale=0.25]{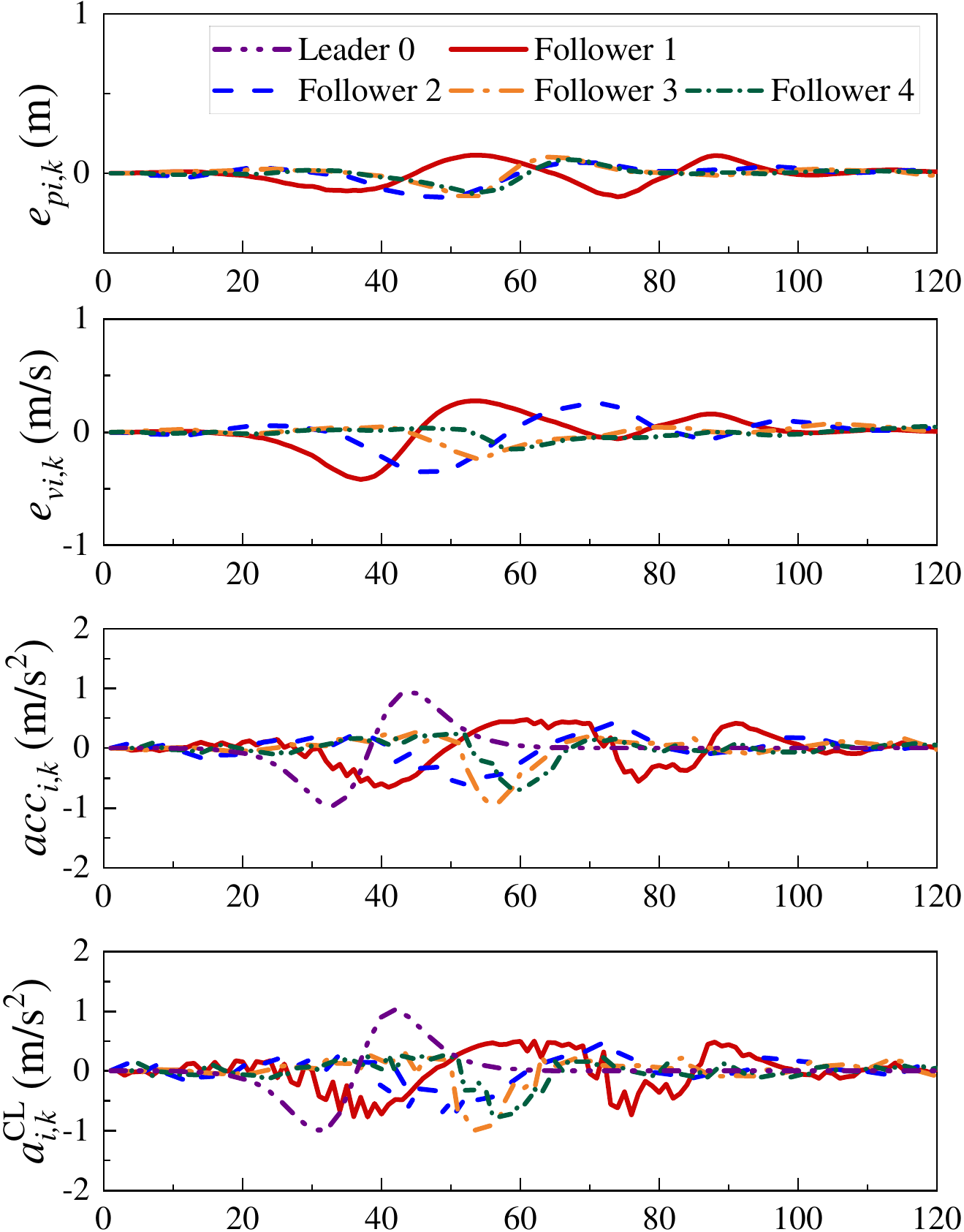}
		\label{RD-PC}
	}
 	\subfigure[PC\_wo\_AS]{
			\includegraphics[scale=0.25]{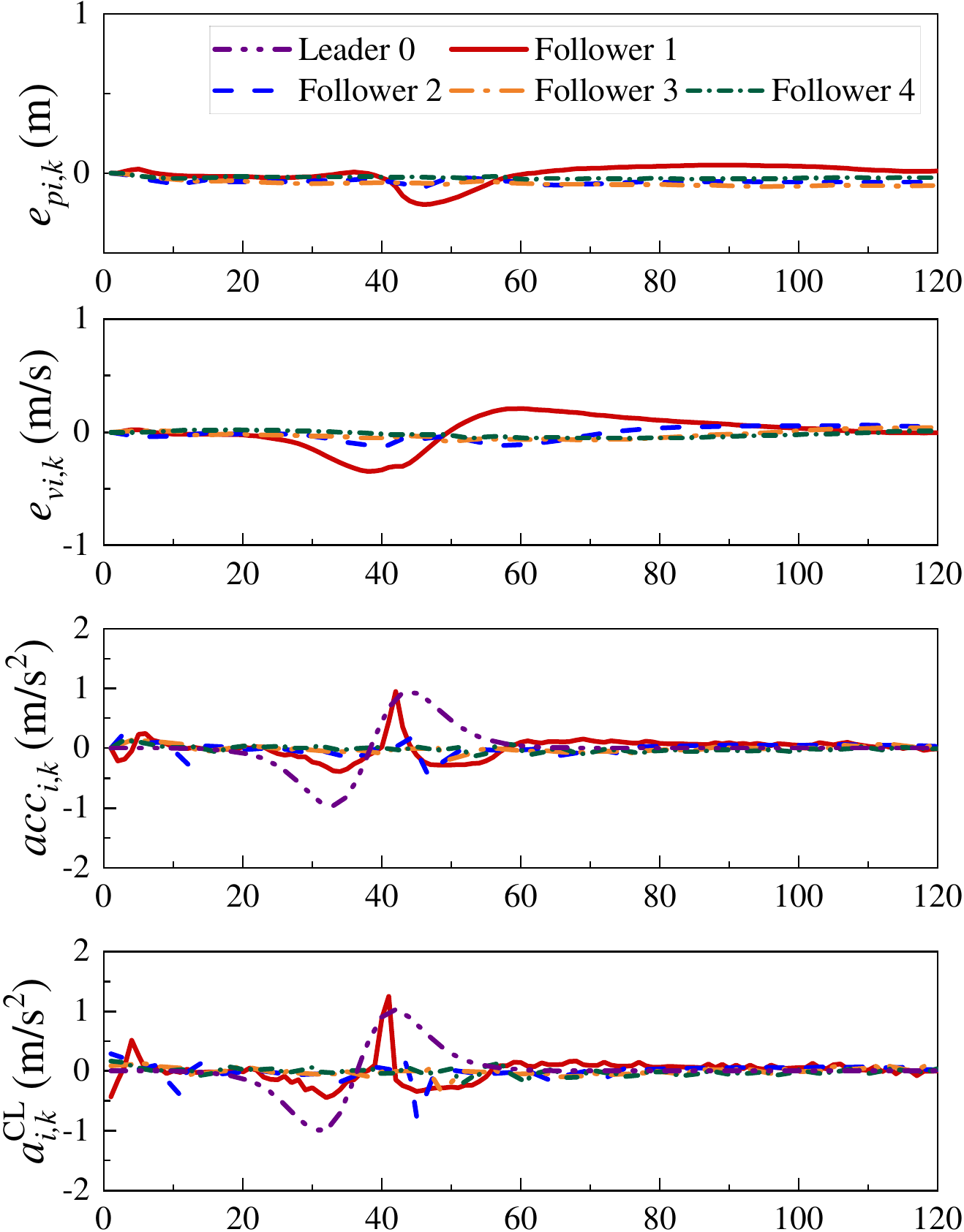}
		\label{PCwoAS}
	}
	\caption{Results of a specific test episode. The driving status $e_{pi,k}$, $e_{vi,k}$, and $acc_{i,k}$ along with the control input $a^{\rm CL}_{i,k}$ and observation delay $\tau_{i,k}$ of each follower $i$ are represented as different curves, respectively.}
	\label{fig_status_Step_1}
\end{figure*}

\subsubsection{Testing results of one episode}
To further examine how the performance improvement of MTCC-PC over the RD-PC and PC\_wo\_AS algorithms in Table~\ref{table_PC_Performance_Step1} is reflected in the physical system, we focus on a specific test episode with $120$ time steps and plot the tracking errors $e_{pi,k}$ and $e_{vi,k}$ of each follower $i\in\mathcal{V}\backslash \{0\}$ as well as the acceleration $acc_{i,k}$ and control input $a^{\rm CL}_{i,k}$ of each vehicle $i\in\mathcal{V}$ for all time steps $k\in\{1,2,\cdots,120\}$. The results for MTCC-PC, RD-PC, and PC\_wo\_AS algorithms are shown in Fig.~\ref{fig_status_Step_1}. \par

Fig.~\ref{fig_status_Step_1} shows that the performance differences among the algorithms are manifested in the speed of convergence to the steady state and the oscillations of the tracking errors, acceleration, and control input. In general, the speed of convergence to the steady state in RD-PC and PC\_wo\_AS for all followers is significantly slower than those in MTCC-PC. Also, the tracking errors, acceleration, and control input in RD-PC and PC\_wo\_AS have larger oscillations than those in MTCC-PC for all followers. 

Specifically, $e_{pi,k}$ for each follower $i\in\mathcal{V}\backslash \{0\}$ in RD-PC reduces to $0$ $\rm m$ (at around $k=100$) later than in MTCC-PC (at around $k=70$). { It can be observed that there are positive gap-keeping errors for followers $1$ in PC\_wo\_AS up to the end of the episode.} The velocity error $e_{vi,k}$ for follower $i$ in RD-PC has larger oscillations than those in MTCC-PC, especially from $k=20$ to $k=110$. { $e_{vi,k}$ of follower $1$ in PC\_wo\_AS has a slower convergence speed to $0$ $\rm m$ than that in MTCC-PC.} Regarding $acc_{i,k}$ and $a^{\rm CL}_{i,k}$, the oscillations in RD-PC are larger than those in MTCC-PC for all followers $i$, especially at $k>20$. In addition, RD-PC also has many more large jerks than MTCC-PC, which greatly reduces driving comfort. { PC\_wo\_AS also has larger jerks than those of MTCC-PC, especially at the beginning of the episode and at around $k=40$.} Although MTCC-PC has a better performance compared to RD-PC and PC\_wo\_AS, there are still many small jerks for $a^{\rm CL}_{i,k}$, especially for followers $1$ and $2$. This is because MTCC-PC is based on a random RRA policy for C-V2X communications. \par

An important requirement for platoon control is to guarantee string stability. When oscillations of the preceding vehicle are attenuated by following vehicles upstream of the platoon, the platoon is considered string stable. For example, as shown in Fig. \ref{MTCC_1}, the amplitudes of the oscillations in $e_{pi,k}$, $e_{vi,k}$, and $acc_{i,k}$ for each follower $i\in\mathcal{V}\backslash \{0\}$ are smaller than those of their respective predecessors $i-1$ in MTCC-PC. The reduction in oscillation amplitude demonstrates the string stability of the platoon. The string stability of the platoon is not satisfactory for RD-PC and PC\_wo\_AS in Fig. \ref{RD-PC} and Fig. \ref{PCwoAS} since the amplitudes of the oscillations in $acc_{i,k}$ of RD-PC for follower $3$ are larger than those for follower $2$ at around $k=58$, { and the amplitudes of the oscillations in $acc_{i,k}$ of PC\_wo\_AS for follower $1$ are larger than those for leading vehicle $0$ at around $k=42$.} {The reason why MTCC-PC performs better in terms of string stability than RD-PC and PC\_wo\_AS is due to the definition of the reward function $R^{\rm CL}_{i,k}$ in \eqref{PC_Reward}. While the first, second, and fourth terms in $R^{\rm CL}_{i,k}$ aim to minimize the absolute value of $e_{pi,k}$, $e_{vi,k}$, and $acc_{i,k}$, the third and fourth terms aim to minimize the value of the control input $a^{\rm CL}_{i,k}$, which will result in smaller oscillations of $e_{pi,k}$, $e_{vi,k}$, and $acc_{i,k}$. Since MTCC-PC achieves better PC performance than RD-PC and PC\_wo\_AS in terms of the expected cumulative reward $J^{\rm CL}_{i}$ in \eqref{pc_perf}, it has a higher probability of satisfying the string stability than RD-PC and PC\_wo\_AS.} \par

\section{Conclusion}
In this paper, we have decomposed the MTCC problem into a communication-aware DRL-based PC sub-problem and a control-aware DRL-based RRA sub-problem. In order to solve the PC sub-problem, we have augmented the PC state space with the observation delay and PC action history, and defined the reward function for augmented state to conceive the augmented state MDP. We have proved that the optimal policy for the MDP is also optimal for the PC problem with observation delay. Finally, the experimental results have demonstrated that (1) training in a delayed environment generated by embedded simulation of C-V2X communications in MTCC rather than by a simple stochastic delay model can improve PC performance, since the delay distribution during training complies with that during execution in practice; { and (2) the PC performance can be improved by augmenting the state with the action history.} In Part II of this { {two-part} paper, we will propose the MTCC-RRA algorithm to learn the RRA policy and design a sample- and computational-efficient training approach to jointly train MTCC-PC and MTCC-RRA algorithms in an iterative process.

\par

\appendix
\subsection{Proof of Theorem 1}
Firstly, we have
\begin{align}
&p(S^{\rm CL}_{i,k+1}|S^{\rm CL}_{i,k},a^{\rm CL}_{i,k})\IEEEnonumber \\
&=p(x_{i,k+1-\tau_{i,k+1}},\{a^{\rm CL}_{i,k'}\}_{k'=k+1-\tau_{\mathrm{max}}}^{k},\tau_{i,k+1}|x_{i,k-\tau_{i,k}},\IEEEnonumber \\
&\{a^{\rm CL}_{i,k'}\}_{k'=k-\tau_{\mathrm{max}}}^{k-1},\tau_{i,k},a^{\rm CL}_{i,k}).
\end{align}\par

According to \eqref{queue}, we discuss the following two situations.

\textbf{1). If $\tau_{i,k+1}=\tau_{i,k}+1$}, we have $x_{i,k+1-\tau_{i,k+1}}=x_{i,k-\tau_{i,k}}$. Therefore,
\begin{align}
&p(S^{\rm CL}_{i,k+1}|S^{\rm CL}_{i,k},a^{\rm CL}_{i,k})\IEEEnonumber \\
&\stackrel{(a)}{=}p(\tau_{i,k+1}|\tau_{i,k},x_{i,k-\tau_{i,k}})\textbf{1}\left\{\{a^{\rm CL}_{i,k'}\}_{k'=k+1-\tau_{\mathrm{max}}}^{k}=\right. \IEEEnonumber \\
&\left. \{a^{\rm CL}_{i,k'}\}_{k'=k+1-\tau_{\mathrm{max}}}^{k-1},a^{\rm CL}_{i,k}\right\}\IEEEnonumber \\
&=p(\tau_{i,k+1}|\tau_{i,k},x_{i,k-\tau_{i,k}}),
\end{align}\par 
\noindent {where the indicator function $\mathbf{1}\{X\}$ is $1$ when $X$ is true and $0$ otherwise.} $p(\tau_{i,k+1}|\tau_{i,k},x_{i,k-\tau_{i,k}})$ in (a) holds since the observation delay $\tau_{i,k}$ is derived from the queue length $q^{\rm CAM}_{i-1,(k-1,T)}$ according to \eqref{eq23}, where $q^{\rm CAM}_{i-1,(k-1,T)}$ is related to transmission rate of V2V link $i-1$ according to \eqref{rateV2VCAM}, which further depends on the observed driving status $x_{i,k-\tau_{i,k}}$.

\textbf{2). If $\tau_{i,k+1}=\tau_{i,k}-d$}, $0\leq d\leq \tau_{i,k}-1$, we have
\begin{align}
&p(S^{\rm CL}_{i,k+1}|S^{\rm CL}_{i,k},a^{\rm CL}_{i,k})\IEEEnonumber \\
&=p(\tau_{i,k+1}|\tau_{i,k},x_{i,k-\tau_{i,k}})\textbf{1}\{\{a^{\rm CL}_{i,k'}\}_{k'=k+1-\tau_{\mathrm{max}}}^{k}=\IEEEnonumber \\
&\{a^{\rm CL}_{i,k'}\}_{k'=k+1-\tau_{\mathrm{max}}}^{k-1},a^{\rm CL}_{i,k}\}\IEEEnonumber \\
&p(x_{i,k+1-\tau_{i,k+1}}|x_{i,k-\tau_{i,k}},\{a^{\rm CL}_{i,k'}\}_{k'=k-\tau_{i,k}}^{k-(\tau_{i,k}+d+1)})\IEEEnonumber \\
&\stackrel{(a)}{=}p(\tau_{i,k+1}|\tau_{i,k},x_{i,k-\tau_{i,k}})\IEEEnonumber \\
&\sum_{x_{i,k-\tau_{i,k}},\dots,x_{i,k-(\tau_{i,k}-d+1)}}\left\{p(x_{i,k-(\tau_{i,k}+1)}|x_{i,k-\tau_{i,k}},a^{\rm CL}_{i,k-\tau_{i,k}})\right.\IEEEnonumber \\
&\left. p(x_{i,k-(\tau_{i,k}+2)}|x_{i,k-(\tau_{i,k}+1)},a^{\rm CL}_{i,k-(\tau_{i,k}+1)})\right.\IEEEnonumber \\
&\left. \dots p(x_{i,k-\tau_{i,k+1}}|x_{i,k-(\tau_{i,k}-d+1)},a^{\rm CL}_{i,k-(\tau_{i,k}-d+1)})\right\},
\end{align}
\noindent where (a) holds since we approximately consider that the $x_{i,k}$ at PC agent $i$ is Markov and therefore the probability of multi-step transition $p(x_{i,k+1-\tau_{i,k+1}}|x_{i,k-\tau_{i,k}},\{a^{\rm CL}_{i,k'}\}_{k'=k-\tau_{i,k}}^{k-(\tau_{i,k}+d+11)})$ is also independent of statue history $\{\dots x_{i,k-\tau_{i,k}-1}\}$. In summary, the above derivation demonstrates that $S^{\rm CL}_{i,k+1}$ only depends on the current state and action pair $\{S^{\rm CL}_{i,k},a^{\rm CL}_{i,k}\}$ but not the history $\{\dots,S^{\rm CL}_{i,k-1}\}$. Therefore, the Markov property is proved for the augmented state $S_{i,k}^{\mathrm{CL}}$.
\subsection{Proof of Theorem 2}
According to \eqref{pc_perf_aug}, we have
\begin{align}
\label{augmentJ}
\tilde{J}^{\rm CL}_{i}&=\mathrm{E}_{\pi^{\rm CM}}\mathrm{E}_{\pi^{\rm CL}_{i}}[\sum_{k=0}^{K-1} \gamma^{k} \tilde{R}^{\rm CL}_{i,k}] \IEEEnonumber \\
&=\mathrm{E}_{\pi^{\rm CM}}\mathrm{E}_{\pi^{\rm CL}_{i}}[\sum_{k=0}^{K-1} \gamma^{k} \mathrm{E}_{x_{i,k}}[R^{\rm CL}_{i,k}(x_{i,k},a^{\rm CL}_{i,k})|S_{i,k}^{\rm CL}]] \IEEEnonumber \\
&=\mathrm{E}_{\pi^{\rm CM}}\mathrm{E}_{\pi^{\rm CL}_{i}}[\sum_{k=0}^{K-1} \gamma^{k} \sum_{x_{i,k}}p(x_{i,k}|S_{i,k}^{\rm CL})[R^{\rm CL}_{i,k}(x_{i,k},a^{\rm CL}_{i,k})]]. \IEEEnonumber \\
\end{align}

In order to prove that $\tilde{J}^{\rm CL}_{i}=J^{\rm CL}_{i}$, and according to the definition $J^{\rm CL}_{i}$ in \eqref{pc_perf}, we must prove that
\begin{align}
\label{eq36}
&\mathrm{E}_{\pi^{\rm CM}}\mathrm{E}_{\pi^{\rm CL}_{i}}[\sum_{k=0}^{K-1} \gamma^{k} \sum_{x_{i,k}}p(x_{i,k}|S_{i,k}^{\rm CL})[R^{\rm CL}_{i,k}(x_{i,k},a^{\rm CL}_{i,k})]] \IEEEnonumber \\
&=\mathrm{E}_{\pi^{\rm CM}}\mathrm{E}_{\pi^{\rm CL}_{i}}[\sum_{k=0}^{K-1} \gamma^{k} R^{\rm CL}_{i,k}(x_{i,k},a^{\rm CL}_{i,k})].
\end{align}

Since both sides of \eqref{eq36} calculate the expected sum of undelayed reward $R^{\rm CL}_{i,k}(x_{i,k},a^{\rm CL}_{i,k})$ between control intervals $[0,K-1]$, the equation holds if the probability distributions of the trajectories $(x_{i,0},a^{\rm CL}_{i,0},\cdots,x_{i,K-1},a^{\rm CL}_{i,K-1})$ are the same on both sides of \eqref{eq36}. Since both sides follow the same PC policy $\pi^{\rm CL}_{i}$ and RRA policy $\pi^{\rm CM}$, we only need to make sure that the distributions of the initial state $x_{i,0}$ are the same on both sides. Note that the distribution of $x_{i,0}$ on the LHS of \eqref{eq36} depends on $S_{i,0}$, i.e., $x_{i,-\tau_{i,0}}$ before control interval $0$. Therefore, given the distribution of $x_{i,0}$, i.e., $p(x_{i,0})$ on the RHS of \eqref{eq36}, if the distribution of $S_{i,0}$, i.e., $p(S_{i,0}^{\rm CL})$ satisfies \eqref{distribution_x} in Theorem 2, the resultant distribution of $x_{i,0}$ on the LHS of \eqref{eq36} is the same as that on the RHS. \par  



Since we have proved that $\tilde{J}^{\rm CL}_{i}=J^{\rm CL}_{i}$, it is obvious that $\tilde{\pi}^{\rm CL*}_{i}=\pi^{\rm CL*}_{i}$ according to \eqref{eq16} and \eqref{optimal_policy_aug}.

\bibliographystyle{IEEEtran}
\bibliography{MTCC}

\end{document}